\documentclass[english,12pt]{article}
\usepackage[a4paper,bottom=4.2cm,top=1cm,head=3cm,width=18cm,dvipdfm]{geometry}
\evensidemargin=\oddsidemargin

\usepackage[T1]{fontenc}
\usepackage{amstext}
\usepackage{babel}
\usepackage{verbatim}
\usepackage{amsfonts}
\usepackage{mathrsfs}
\usepackage{amsmath}
\usepackage{amssymb}
\usepackage{bm}
\usepackage{extarrows}
\usepackage{slashed}
\usepackage{isodateo}
\usepackage{xcolor}
\usepackage{graphicx}
\usepackage[low-sup]{subdepth}
\usepackage{isodateo}
\usepackage{caption}
\usepackage{subcaption}
\usepackage[linktocpage=true]{hyperref}
\hypersetup{
	colorlinks=true,%
	linkcolor=[rgb]{0,0,1}, 
	citecolor=[rgb]{0,0,1}
}

\numberwithin{equation}{section}

\makeatletter
\@ifundefined{textcolor}{}
{%
	\definecolor{BLACK}{gray}{0}
	\definecolor{WHITE}{gray}{1}
	\definecolor{RED}{rgb}{1,0,0}
	\definecolor{GREEN}{rgb}{0,1,0}
	\definecolor{BLUE}{rgb}{0,0,1}
	\definecolor{CYAN}{cmyk}{1,0,0,0}
	\definecolor{MAGENTA}{cmyk}{0,1,0,0}
	\definecolor{YELLOW}{cmyk}{0,0,1,0}
}
\newcommand{\fr}[2]{\mbox{$\frac{\,{#1}\,}{#2}$}}
\renewcommand{\rm}{\mathrm}
\def\bge{\begin{equation}}
	\def\ede{\end{equation}}
\def\bga{\begin{aligned}}
	\def\eda{\end{aligned}}
\newcommand{\beq}{\begin{equation}}
	\newcommand{\eeq}{\end{equation}}
\newcommand{\bq}{\begin{equation}}
	\newcommand{\eq}{\end{equation}}
\newcommand{\ba}{\begin{array}}
	\newcommand{\ea}{\end{array}}
\newcommand{\beqa}{\begin{eqnarray}}
	\newcommand{\eeqa}{\end{eqnarray}}
\newcommand{\beqs}{\begin{subequations}}
	\newcommand{\eeqs}{\end{subequations}}

\def\nn{\nonumber}

\def\({\left(}
\def\){\right)}

\def\leqq{\leqslant}

\def\End{\end{document}}

\def\al{\alpha}

\def\ga{\gamma}

\renewcommand{\rm}{\mathrm}
\def\bge{\begin{equation}}
\def\ede{\end{equation}}
\def\bga{\begin{aligned}}
\def\eda{\end{aligned}}

\def\nn{\nonumber}

\def\({\left(}
\def\){\right)}
\def\[{\left[}
\def\]{\right]}

\def\End{\end{document}}

\def\leqq{\leqslant}

\def\al{\alpha}

\def\ga{\gamma}

\def\M{\mathcal{M}}

\def\to{\rightarrow}

\def\AA{\mathcal{A}}

\def\O{\mathcal{O}}

\def\hs{\hspace*{0.3mm}}

\def\hsm{\hspace*{-0.3mm}}
\def\hsmx{\hspace*{-0.5mm}}

\def\to{\rightarrow}

\newlength{\halfpagewidth}
\divide\halfpagewidth by 2

\def\End{\end{document}}
\begin{document}

\vspace{1cm}
\begin{center}
	{\Large\bf UV Completion of Neutral Triple Gauge Couplings}
	\vspace*{1cm}
	
{\bf John Ellis}$^{1,2,3}$, 
~{\bf Hong-Jian He}$^{1,4,5}$, 
~{\bf Rui-Qing Xiao}$^{1,2,4}$, 
\\
{\bf Shi-Ping Zeng}$^{1,4}$,
~{\bf Jiaming Zheng}$^{1,4}$
	
\vspace*{3mm}
{\small 
$^1$\,{T.~D.~Lee Institute, Shanghai Jiao Tong University, Shanghai, China} \\  
$^2$\,{Department of Physics, King's College London, Strand, London WC2R 2LS, UK}\\
$^3$\,{Theoretical Physics Department, CERN, CH-1211 Geneva 23, Switzerland}  \\
$^4$\,{School of Physics and Astronomy, 
Key Laboratory for Particle Astrophysics and Cosmology, 
Shanghai Key Laboratory for Particle Physics and Cosmology, \\
Shanghai Jiao Tong University, Shanghai, China} \\
$^5$\,Physics Department, Tsinghua University, Beijing, China; \\
      Center for High Energy Physics, Peking University,  Beijing, China  
\\[5mm]
{\tt john.ellis@cern.ch, hjhe@sjtu.edu.cn, xrq12@tsinghua.org.cn,\\ spzeng@sjtu.edu.cn, zhengjm3@gmail.com}
}
\end{center}

\vspace{5mm}
\begin{abstract}
\vspace{5mm}
\noindent 
Neutral triple gauge couplings (nTGCs) are manifestation of new physics beyond the Standard Model (SM),
as they are absent in the SM and are first generated by dimension-8 operators
in the SM Effective Field Theory (SMEFT).\
We study the UV completion of nTGCs in a renormalizable model with vector-like heavy fermions.\
We compute the one-loop heavy fermion contributions to nTGC vertices by matching them to dimension-8 operators
in the low energy limit.\ Such fermion loops contain either heavy fermions only or mixture of heavy fermions
with light SM fermions. We find that their contributions can induce dimension-8 nTGC effective operators 
containing two SM Higgs-doublet fields, which are formulated with a complete set of 7 dimension-8 operators
generating off-shell CP-even nTGCs.\
We present the results in terms of SMEFT coefficients and in terms of nTGC vertices
(form factors) with two on-shell gauge bosons.\  In the heavy-light mixing case there appear terms that
cannot be accommodated by conventional parametrizations of form factors due to extra logarithmic corrections.\
We further discuss the implications for probing such UV dynamics via nTGCs at high-energy colliders.
\\[1cm]
KCL-PH-TH/2024-45, CERN-TH-2024-143 
\\[2mm]
{Phys.\ Rev.\ D (2024) in Press [\,arXiv:2408.12508\,]}
\end{abstract}

\newpage
\baselineskip 18pt
\tableofcontents{}

\vspace*{5mm}

\section{\large\hspace*{-2.5mm}Introduction}
\label{sec:1}

Neutral Triple Gauge Couplings\,(nTGCs) are sensitive probes of new physics beyond the Standard Model (SM)
because they are absent in the SM and first show up in the SM Effective Field Theory
(SMEFT)\,\cite{Buchmuller:1985jz}\cite{Grzadkowski:2010es}\cite{SMEFT}
as manifestations of dimension-8 operators.\
For these reasons, they have been subject to experimental searches
by the ATLAS\,\cite{ATLAS:2018nci} and CMS collaborations\,\cite{CMS:2016cbq}, 
and have recently attracted widespread phenomenological interest\,\cite{Ellis:2023ucy}-\cite{nTGC-other}.\
In most of these studies, the nTGC signals would appear in the production of
two on-shell neutral bosons $Z\gamma$ or $ZZ$ via an $s$-channel virtual neutral vector boson.$\!$\footnote{%
The general formulation of the nTGC vertices and form factors with two off-shell vector bosons as well as 
its important application to analyzing the LHC production of $Z^*\gamma\,(\nu\bar{\nu}\ga)$ 
was presented in Ref.\,\cite{Ellis:2023ucy}.}\
When the momentum dependence of the vertex is polynomial,
as when it is generated by tree-level contributions from effective operators,
it is a convenient practice to enumerate the relevant tensor structures and associated
form factors of the vertices with one off-shell and two on-shell neutral gauge bosons.\
Up to cubic dependence on the particle momenta, there are 6 CP-conserving tensor structures
for all possible combinations of triple gauge boson vertices\,\cite{Hagiwara:1986vm}.

\vspace*{1mm}

The parametrization of nTGCs in the framework of the
SMEFT operators\,\cite{Ellis:2023ucy}-\cite{Ellis:2020ljj}\cite{Degrande:2013kka}
has several benefits over the conventional form factor formulation.\
Most notably, it maintains the SM gauge symmetry manifestly,
which is essential to eliminate unphysical energy dependences as required by the SM 
with spontaneous electroweak gauge symmetry breaking\,\cite{Ellis:2023ucy}\cite{Ellis:2022zdw}.\
It also provides a general framework for studying processes with one or more
off-shell gauge bosons in the nTGC vertex.\
The effective operators that generate nTGCs first appear at the dimension-8 level of the SMEFT,
and have been the starting point of many recent phenomenological studies\,\cite{Ellis:2023ucy}-\cite{nTGC-other}.\
Computationally, the effective field theory (EFT) approach has the benefit of separating clearly 
the UV and IR contributions from the underlying physics\,\cite{Georgi:1993mps},
a feature that we use extensively in this work.

\vspace*{0.6mm}

An EFT analysis is general only when all the operators satisfy the assumed symmetry under consideration.\
But, such an analysis could become cumbersome and non-intuitive if the number of contributing
operators is large.\ The common compromise such as operator-by-operator analysis
trades generality for simplicity of the analysis.\
Because of the freedom in the choice of operator basis,
a simple UV model does not necessarily correspond to a small number of effective operators
in the IR unless the symmetry of the UV theory restricts it tightly.\
Moreover, in a given UV model various operators may be generated at different loop orders 
and the coefficients of the operators may have different magnitudes from the naive dimensional
power counting, so it is important to analyze explicitly certain UV models as benchmarks 
and understand their low-energy contributions to the corresponding SMEFT operators.

\vspace*{0.6mm}

Previous literature on the UV origin of nTGC vertices conventionally focused on U(1)-invariant
form factors\,\cite{Gounaris:2000tb}\cite{Gounaris:2000dn}
rather than SMEFT operators.\
The main purpose of this work is to explore how CP-conserving nTGC operators that generate SM SU(2)$\otimes$U(1) form factors
can be generated from the underlying renormalizable and perturbative UV models.\
We demonstrate that the dimension-8 nTGC operators induced by fermionic one-loop contributions
must contain two Higgs-doublet fields, and that the dimension-8 higgsless (pure gauge) operators for nTGCs cannot be generated in this way.\
The fermionic UV models we study either contain two new heavy-fermion multiplets
that couple to the SM Higgs field through Yukawa-like couplings (the ``all-heavy'' case),
or contain a single heavy-fermion multiplet that couples to the SM chiral fermions via a Higgs doublet
(the ``heavy-light'' case).\ In order to match the nTGC vertices in the UV models
to those of the low-enenrgy effective theory, we compute the loop diagrams using the method of regions\,\cite{Georgi:1993mps}\cite{Chetyrkin:1988zz}-\cite{Semenova:2018cwy}
that separates the contributions from loop momenta in the IR and UV regions.\
The former (soft part) matches the light-fermion-loop diagram of tree-level effective operators,
whereas the latter (hard part) directly matches the heavy-fermion-loop-induced effective operator in the SMEFT.\
A nontrivial technical issue concerns the treatment of the mixed loop diagrams
that contain both the light SM chiral fermions and the new heavy fermions.\

Another technical issue arising in the computation of the fermion loop diagrams is the ambiguity of the $\gamma_{5}$
definition\,\cite{Jegerlehner:2000dz} in dimensional regularization (DREG).\
Here we adopt the naive dimensional regularization (NDR)
scheme\,\cite{Kreimer:1989ke}-\cite{Kreimer:1993bh}
that maintains the anti-commutativity of the $\gamma$ matrices in $D$ dimensions
and has been shown to preserve gauge invariance automatically
in the renormalization of loop diagrams.\
This is more convenient than non-anticommuting schemes
such as the Breitenlohner-Maison-'t\,Hooft-Veltman (BMHV)
scheme\,\cite{tHooft:1972tcz}-\cite{Belusca-Maito:2020ala},
where gauge invariance is imposed manually by adding finite counter terms.\
Nevertheless, in the context of EFT matching, the soft and hard parts of the loop diagrams
may contain canceling finite terms that violate the gauge invariance of each part separately.\
This is closely related to the irrelevant anomalies in the EFT calculation discussed in the
recent literature\,\cite{Feruglio:2020kfq}-\cite{Cohen:2023gap}.
In order to circumvent the need to introduce finite counter terms,
we discuss carefully the Ward-Takahashi identity associated with the loop diagrams under consideration and prescribe a rule for choosing reading points of spinor traces in NDR that eliminates the appearance of irrelevant
anomalies in all the intermediate steps of matching.

\vspace*{0.6mm}

Our calculations yield comparable nTGC vertices for the ``all-heavy'' and ``heavy-light'' scenarios.\
The familiar perturbative loop factors reduce the values of coefficients of
the corresponding dimension-8 SMEFT operators to be smaller than
what might be expected from naive dimensional analysis.\
We compare the sensitivities of collider probes of nTGCs estimated 
by the recent phenomenological studies\,\cite{Ellis:2023ucy}-\cite{Liu:2024tcz}
with the contributions of the heavy fermion loops, 
discussing the prospects for direct confrontations between direct and indirect searches for such new physics.\
Observation of some nTGCs without the corresponding discovery of new heavy fermion
would suggest that the nTGCs originate from strong dynamics beyond the SM.\

\vspace*{0.8mm}

The layout of this paper is as follows.\ 
In Section\,\ref{sec:2}
we give the complete set of seven dimension-8 SMEFT operators
that contribute to nTGCs, and their matching with one-loop perturbative calculations is
studied in subsequent sections.\
In Section\,\ref{sec:3}
we discuss the general structure of heavy-fermion one-loop diagrams that contribute to the nTGCs.\
Section\,\ref{sec:4} describes our calculational method of momentum integration by regions
and the matching to the coefficients of dimension-8 SMEFT operators that contribute to nTGCs,
where our treatment of $\gamma_5^{}$ in diagrams involving heavy-light mixing is discussed in detail.\
We present in Section\,\ref{sec:5} our results for the contributions to nTGCs from loops 
with heavy fermions only and from loops with heavy-light mixing.\
Finally, we draw conclusions from this study in Section\,\ref{sec:6}.

\section{\large\hspace*{-2.5mm}CP-Conserving nTGC Operators of Dimension-8}
\label{sec:general_operators}
\label{sec:2}

In this Section we present the complete set of dimension-8 nTGC operators in the SMEFT, 
which are needed for matching with the perturbative one-loop contributions 
of the UV completion model.\
These operators contain terms with three neutral gauge bosons 
and a number of Higgs fields that acquire expectation values in the symmetry breaking phase.\ 
At this level, the renormalizable fermionic model that we consider as the UV completion 
can only induce nTGC SMEFT operators that contain Higgs-doublet fields.\ 
To make this clear, we render manifest the SU(2)$\otimes$U(1) electroweak gauge symmetry  
of the SMEFT by working in the symmetric phase, so that all particles appearing in the loop 
are gauge multiplets of SU(2)$\otimes$U(1).\ 
It is then easy to see that loop diagrams with only SU(2)$\otimes$U(1)-invariant pure gauge vertices 
do not induce nTGC interaction at one-loop order.\ 
This is because, in the symmetric phase, the SM gauge interactions do not mix chiral fermions 
with new massive fermions.\ 
Any fermion loop diagram with only fermion-gauge vertices must 
either contain massless SM chiral fermions only or heavy vector fermions only.\ 
However, loop diagrams with only {massless} SM chiral fermions do not 
induce effective operators of the SMEFT, which arise from integrating out heavy particles in the UV theory.\ 
We note also that such loops with three external gauge bosons  
are also constrained by gauge anomaly cancellation conditions and so cannot contribute to nTGC vertices.\ 
On the other hand, in the absence of chiral fermions, the gauge interactions preserve 
charge conjugation ($C$) symmetry with $C$-odd vector bosons.\
In this case, triple gauge boson couplings violate charge conjugation, 
and thus nTGCs cannot be generated by loop diagrams with only gauge vertices that preserve charge conjugation. 
Thus, the loop diagrams should contain other $C$-violating sources such as Yukawa couplings 
to the SM Higgs doublet or Yukawa-like couplings to certain new heavy scalar fields.\  
Hence, the SMEFT nTGC operators containing pure gauge fields alone 
can only arise from contracting the additional fields (such as the new heavy scalars) 
with $C$-violating vertices in the loop which should be at least of two-loop order.\ 
Such Higgsless contributions should be suppressed 
by the two-loop factors unless the UV theory is strongly-coupled 
and generates the nTGC operators  non-perturbatively\,\cite{Gounaris:2000tb}.\ 
Such a strongly-interacting UV theory is an interesting possibility,  
but is beyond the scope of this study.\ 
For the present work, we focus on a perturbatively renormalizable UV theory 
including vector-like new heavy fermions, whose one-loop contributions can induce 
the dimension-8 nTGC operators containing Higgs-doublet fields in the low-energy SMEFT.\

\vspace*{0.6mm}

There are 7 independent CP-conserving nTGC operators with two SM-Higgs-doublet fields 
after accounting for the equivalence due to integration by parts.\ 
We choose the following operator basis for our nTGC analysis:
\\[-8mm]
\begin{subequations}
\begin{align}
\O_{\tilde{W}W}^{} &= 
c_{\tilde{W}W}^{}\hsm 
\big[ \text{i}H^{\dagger}\tilde{W}_{\mu\nu}W^{\nu\rho}\{D_{\rho},D_{\mu}\}H + \text{h.c.}
\big]  , \\
\O_{\tilde{W}W}^{\prime} & =  
c_{\tilde{W}W}^{\prime}
\hsm\big[
\text{i}H^{\dagger}\tilde{W}_{\mu\nu}(D_{\rho}W^{\nu\rho})D_{\mu}H + \text{h.c.} \big], 
\\
\O_{\tilde{B}B} &= 
c_{\tilde{B}B}
\hsm\big[ 
\text{i}H^{\dagger}\tilde{B}_{\mu\nu}B^{\nu\rho}\{D_{\rho},D_{\mu}\}H + \text{h.c.}
\big], \\
\O_{\tilde{B}B}^{\prime} &=
c_{\tilde{B}B}^{\prime}
\hsm\big[ 
\text{i}H^{\dagger}\tilde{B}_{\mu\nu}(D_{\rho}B^{\nu\rho})D_{\mu} H + \text{h.c.}
\big],  
\end{align}
\label{eq:operators1} 
\vspace*{-4.5mm}
\end{subequations}
and
\begin{subequations}
\vspace*{-1.5mm}
\begin{align}
\O_{\tilde{B}W} &= 
c_{\tilde{B}W}
\hsm\big[ 
\text{i}H^{\dagger}\tilde{B}_{\mu\nu}W^{\nu\rho}\{D_{\rho},D^{\mu}\}H + \text{h.c.}
\big] , \\
\O_{\tilde{B}W}^{\prime} &= 
c_{\tilde{B}W}^{\prime}
\hsm\big[ 
\text{i}H^{\dagger}\tilde{B}_{\mu\nu}(D_{\rho}W^{\nu\rho})D^{\mu}H + \text{h.c.}
\big], \\
\O_{\tilde{W}B} &= 
c_{\tilde{W}B}
\hsm\big[ 
\text{i}H^{\dagger}\tilde{W}_{\mu\nu}B^{\nu\rho}\{D_{\rho},D^{\mu}\}H + \text{h.c.} 
\big] ,
\end{align}
\label{eq:operators2} 
\end{subequations}
\hspace*{-3mm}
where we use the notations $W_{\mu\nu}\!=\! W_{\mu\nu}^{I}\sigma^{I}\!/2$ and  $\tilde{F}_{\mu\nu}\!=\!\frac{1}{2}\epsilon_{\mu\nu\rho\sigma}F^{\rho\sigma}$,
and denote the SM-Higgs-doublet field by $H$ with vacuum expectation value (VEV)
$H_0^{}\!=\!\left<H\right>$.\ 
The coefficient of each nTGC operator above is related to the UV cutoff scale $\Lambda$ 
of the SMEFT as follows:
\begin{equation}
    c_i^{}\,\equiv\,\frac{\,\bar{c}_i^{}\,}{\,\Lambda^4\,}\,,
\end{equation}
and $\bar{c}_i^{}$ is the corresponding dimensionless coupling coefficient.\
The Jacobi identity implies:
\beq 
\frac{1}{2}(D_{\mu}\tilde{F}_{\gamma\delta})F'^{\gamma\delta}
=\frac{1}{2}(D_{\mu}F_{\gamma\delta})\tilde{F}'^{\gamma\delta}=(D_{\alpha}F_{\beta\mu})\tilde{F}'^{\alpha\beta} \, .
\eeq
Hence, operators of the type 
$\text{i}H^{\dagger}\hsm\tilde{F}_{\mu\nu}F^{\mu\nu}\!D^2\hsm H$ or 
$\text{i}H^{\dagger}\tilde{F'}_{\mu\nu}(D_{\rho}F^{\mu\nu})D^{\rho}\hsm H$ 
can be converted to those in Eq.\eqref{eq:operators2}, 
up to terms that do not contribute to nTGCs.\ 
The seven operators in Eqs.\eqref{eq:operators1}
and \eqref{eq:operators2} are general in the sense 
that every dimension-8 SMEFT operator that contributes to nTGC with two Higgs-doublet fields 
can be reduced to linear combinations of these operators plus terms that are irrelevant for nTGCs, 
up to integration by parts or the Schouten identity,
\begin{equation}
g_{\mu\nu}^{}\epsilon_{\alpha\beta\gamma\delta}^{}
+g_{\mu\alpha}^{}\epsilon_{\beta\gamma\delta\nu}^{}
+g_{\mu\beta}^{}\epsilon_{\gamma\delta\nu\alpha}^{}
+g_{\mu\gamma}^{}\epsilon_{\delta\nu\alpha\beta}^{}
+g_{\mu\delta}^{}\epsilon_{\nu\alpha\beta\gamma}^{} =0\,.
\label{eq:Schouten}
\end{equation}
The completeness of these operators can be verified by counting the number of independent tensor structures 
of the off-shell vertices that contain three powers of momenta, an anti-symmetric tensor, 
and three non-contracted indices from external gauge bosons.\ 
After accounting for bosonic symmetry, the triple gauge boson vertex 
$W^{\mu}(-p_{1}-p_{2})$-$W^{\nu}(p_{1})$-$W^{\rho}(p_{2})$ 
 has only two independent Lorentz structures:
\begin{subequations}
\begin{align}
 & \big[\!-p_{2}^{2}\hs p_{1\sigma}^{} \!+\hsm p_{1}^{2}\hs p_{2\sigma}^{}
   \!-\hsm\!2(p_{1}^{}\!\cdot\hsm p_{2}^{})(p_{2}^{}\!-p_{1}^{})_{\sigma}
   \!-\!2p_{2}^{2}\hs p_{2\sigma}^{}\!+\!2p_{1}^{2}\hs p_{1\sigma}^{}\big]
 \epsilon^{\mu\nu\rho\sigma} , 
 \\[1mm]
 & \big[(p_{1}^{}\!+\hsm p_{2}^{})^{\mu}\epsilon^{\nu\rho\alpha\beta}
 \!+p_{1}^{\nu}\epsilon^{\mu\rho\alpha\beta}
 \!-p_{2}^{\rho}\hs\epsilon^{\mu\nu\alpha\beta}\big] p_{2\alpha}^{}\hs p_{1\beta}^{} \hs ,
\end{align}
\end{subequations}
which correspond to linear combinations of $\O_{\tilde{W}W}$ and
$\O_{\tilde{W}W}^{\prime}$.\ 
For $W^{\mu}(-p_{1}-p_{2})$-$B^{\nu}(p_{1})$-$B^{\rho}(p_{2})$ vertices, 
the Bose symmetry and Schouten identity enforce:
\begin{equation}
0=
\(-p_{2}^{\mu}\hs\epsilon^{\nu\rho\alpha\beta}\!+p_{2}^{\nu}\hs\epsilon^{\mu\rho\alpha\beta}
\!-p_{2}^{\rho}\hs\epsilon^{\mu\nu\alpha\beta}\)\!p_{2\alpha}^{}\hs p_{1\beta}^{}
\!+\!(p_{2}^{2}\hs p_{1\sigma}^{}\!-p_{1}^{}\!\cdot\hsm p_{2}^{}\, p_{2\sigma}^{})
\epsilon^{\mu\nu\rho\sigma}\hsm\!+\! 
(p_{1}\!\leftrightarrow\! p_{2}, \nu\!\leftrightarrow\!\rho)\hs .
\label{eq:schouten_pp}
\end{equation}
Hence the $W$-$B$-$B$ vertex has five independent structures:
\begin{subequations}
\begin{align}
 & (p_{1}^{2}\hs p_{1\sigma}^{}\!-\hsm p_{2}^{2}\hs p_{2\sigma}^{})\hs\epsilon^{\mu\nu\rho\sigma}\,,
 \\
 & (p_{1}^{}\!-\hsm p_{2}^{})^{\mu}\epsilon^{\nu\rho\alpha\beta}p_{2\alpha}^{}\hs p_{1\beta}^{}\,,
 \\
 & (p_{1}^{\nu}\hs\epsilon^{\mu\rho\alpha\beta}\!-\hsm p_{2}^{\rho}\hs\epsilon^{\mu\nu\alpha\beta})
 \hs p_{2\alpha}^{}\hs p_{1\beta}^{}\,,
 \\
 & (p_{1}^{\rho}\epsilon^{\mu\nu\alpha\beta}\!-\hsm p_{2}^{\nu}\epsilon^{\mu\rho\alpha\beta})\hs 
 p_{2\alpha}^{}\hs p_{1\beta}^{}\,,
 \\
 & (p_{1}^{}\!\cdot\hsm p_{2}^{})\hs\epsilon^{\mu\nu\rho\sigma}(p_{2\sigma}^{}\!-\hsm p_{1\sigma}^{})\hs,
\end{align}
\end{subequations}
which do not receive any contributions from $\O_{\tilde{W}W}$ and
$\O_{\tilde{W}W}^{\prime}$, and thus correspond to five more independent
operators.\ These together account for all the seven independent operators 
in Eqs.\eqref{eq:operators1} and \eqref{eq:operators2}.\ 
By inspection, we find that they also span
the 7 different tensor structures of the $B$-$B$-$B$ and $B$-$W$-$W$ vertices.\ 
Hence, these $7$ operators form a complete basis for the tensor structures 
of nTGC vertices including two Higgs-doublet fields.\ 
These operators are linear combinations of 
$\big({\cal O}_{W^{2}\phi^{2}D^{2}}^{(9)}$,
${\cal O}_{W^{2}\phi^{2}D^{2}}^{(17)}$, 
${\cal O}_{WB\phi^{2}D^{2}}^{(14)}$,
${\cal O}_{WB\phi^{2}D^{2}}^{(15)}$, 
${\cal O}_{WB\phi^{2}D^{2}}^{(18)}$,
${\cal O}_{B^{2}\phi^{2}D^{2}}^{(10)}$, 
${\cal O}_{B^{2}\phi^{2}D^{2}}^{(12)}\big)$
listed in Table\,2 of the dimension-8 SMEFT analysis in \cite{Chala:2021cgt}, up to non-nTGC terms.%
\footnote{ See also \cite{Ren:2022tvi} for a complete set of the dimension-8 SMEFT
 operators in the off-shell Green's basis.}

\vspace*{0.6mm}

In general, equations of motions~(EOMs) can be used to convert some of these operators to operators with currents but no explicit TGC  structure, as was done in \cite{Degrande:2013kka}.\ 
Using this procedure, one could reduce the set of seven operators to just one remaining operator with explicit nTGC structure.\ 
But, for this study we work with the complete set of seven nTGC operators 
in Eqs.\eqref{eq:operators1}-\eqref{eq:operators2} 
for the following reasons.\ 
(i).\,It is more convenient to perform SMEFT matching with 
the off-shell nTGC diagrams in such a basis 
because we only need to consider diagrams with external gauge bosons 
rather than fermions and scalar fields in the currents.\  
(ii).\,While it is possible to use EOMs to relate the two nTGC operators $\O_{G1,2}^{}$
and operators with currents (denoted by $\O_{C}^{i}$), 
i.e., $\O_{G1}^{}\!=a_{G2}^{}\O_{G2}^{}\hsm + \sum a_{C}^i \O_{C}^{i}\hsm +\cdots$,
the converted combination of operators $\sum a_{C}^i \O_{C}^{i}$
still describe part of the UV physics.%
\footnote{Here the $\O_{C}^{i}$ may not include the full set of operators 
with fermionic currents in a specific EFT basis.}\
Hence, for the present study we include both $O_{G1}^{}$ and $O_{G2}^{}$ 
rather than only one of them in the matching procedure;
so the triple gauge boson vertices induced by the UV theory are described by the nTGC operators 
rather than those involving fermion currents in the EFT.\ 
(iii).\,In a general SMEFT analysis, one needs to include {\em all} the operators of 
the given order that satisfy the SM electroweak gauge symmetry SU(2)$\otimes$U(1), 
without making further assumptions on the underlying UV physics.\ 
We consider in this work the UV physics that generates the neutral triple gauge boson couplings.\ 
From the IR point of view, the complete prediction for a process,
such as $\bar{f}f\!\to\! VV\,$ induced by nTGCs 
should include either the contributions from all the seven nTGC operators listed above,%
\footnote{For a given UV model, additional effective operators other than nTGC operators, 
such as those involving SM fermions (e.g., the contact operators of $f\bar{f}VV$), 
may contribute to the $\bar{f}f\!\to\!VV$ process in the IR limit.}\
or a single nTGC operator and all other operators with currents that are obtained by using the EOMs.\ 
Here we choose to work with all the seven nTGC operators, 
anticipating that a given UV theory would contribute to the coefficients of
most of these operators.\ Restricting to only a few operators in the SMEFT is not justified 
before knowing the low-energy predictions of a given UV theory.

\vspace*{0.6mm}

The nTGCs can be formulated through effective vertices of the types $V^{*}Z\gamma$
and $V^{*}ZZ$, where $V^{*}$ denotes a virtual $Z^*$ or $\gamma^*$ gauge boson.\ 
Conventionally,  the nTGC vertices can be parametrized 
as follows\,\cite{Gaemers:1978hg}\cite{Hagiwara:1986vm}\cite{Degrande:2013kka}\cite{Ellis:2022zdw}\cite{Ellis:2023ucy}:
\begin{subequations}
\begin{align}
\Gamma_{V^{*}\gamma Z}^{\mu\nu\alpha}(q,p_{1}^{},p_{2}) 
& = \frac{\,c_{V^*\gamma Z}^{}\,}{\,m_Z^2\,} 
(q^{2}\!- m_{V}^{2}) p_{1\beta}^{}\hs\epsilon^{\mu\nu\alpha\beta} ,
\\[0.5mm]
\Gamma_{V^{*}ZZ}^{\mu\nu\alpha} (q,p_{1}^{},p_{2}^{}) 
& = \frac{\,c_{V^*ZZ}^{}\,}{\,m_Z^2\,}
(q^{2}\!-m_{V}^{2})\!
\(p_{1}^{}\!-p_{2}^{}\)_{\beta}^{}
\epsilon^{\mu\nu\alpha\beta} ,
\end{align}
\label{eq:on-shell_vertices} 
\end{subequations}
\hspace*{-2mm}
which contribute to the simplest production process  $f\bar{f}\!\to\! V_{1}V_{2}\,$.\ 
The above nTGC form factor coefficients $(c_{V^*\gamma Z}^{},\hs c_{V^*ZZ}^{})$
are connected to the conventional $(h_3^V,\hs f_5^V)$ notation for nTGC form factors
\cite{Degrande:2013kka}\cite{Ellis:2022zdw}\cite{Ellis:2023ucy} 
via the relation:
\\[-6mm]
\beq 
\label{eq:CVVZ-h3f5} 
(c_{V^*\gamma Z}^{},\hs c_{V^*ZZ}^{}) = (e h_3^V,\hs e f_5^V)\hs, 
\eeq 
where $e$ is the electric charge.\ 
These expressions are enforced by bosonic symmetry, the gauge invariance of the photonic interactions, the on-shell condition of the external vector bosons $\ga Z$ and $ZZ$, and the assumption of cubic dependence on external momenta.\
We have also neglected in (\ref{eq:on-shell_vertices}) any terms
that are proportional to $q^{\mu}_{V}$, since in collider processes
such as $f\bar{f}\!\to\! V_{1}V_{2}$, they will be contracted with
the on-shell fermion current and thus their contributions to production amplitudes
are suppressed by the negligible light fermion mass $m_f^{}\hs$.\ 
In this approximation, the vertices with two on-shell photons vanish. 

\vspace*{0.6mm}

The nTGC operators \eqref{eq:operators1} and \eqref{eq:operators2}
correspond to the hard parts of the loops in the UV theory.\ 
Their contributions to the coefficients are given by
\begin{subequations}
\begin{align}
\Delta c_{\gamma^{*}\gamma Z}^{} & 
=\fr{1}{4}m_{Z}^{3}v\big[\!-\sin(2\theta_{W}^{}\hsmx )c_{\tilde{B}W}'\!+\hsm 4\cos^{2}\!\theta_{W}^{}c_{\tilde{B}B}'
  \!+\hsm \sin^{2}\! \theta_{W}^{}c_{\tilde{W}W}'\big],
\\[0.5mm]
\Delta c_{Z^{*}\gamma Z}^{} & 
=\fr{1}{8}m_{Z}^{3}v\big[4c_{\tilde{B}W}^{}\!-\hsm 4c_{\tilde{W}B}^{}\!-\hsm 4\cos^2\!\theta_W^{}c_{\tilde{B}W}'
 \!-\hsm 4\sin(2\theta_{W}^{})c_{\tilde{B}B}'\!+\hsm\sin(2\theta_{W}^{}\hsmx )c_{\tilde{W}W}'\big],
\\[0.5mm]
\Delta c_{\gamma^{*}ZZ}^{} & 
=\fr{1}{8}m_Z^{3}v\big[\!-4c_{\tilde{B}W}^{}\!+\hsm 4c_{\tilde{W}B}^{}\!+\hsm 4\sin^2\!\theta_{W}^{}c_{\tilde{B}W}'
 \!-\hsm 4\sin(2\theta_W^{})c_{\tilde{B}B}'\!+\hsm\sin(2\theta_W^{}\hsmx )c_{\tilde{W}W}'\big] ,
\\[0.5mm]
\Delta c_{Z^{*}ZZ}^{} & 
=\fr{1}{4}m_{Z}^{3}v\big[\!\sin(2\theta_W^{}\hsmx )c_{\tilde{B}W}'\!+\hsm 
4\sin^2\!\theta_{W}^{}c_{\tilde{B}B}' \!+\hsm \cos^{2}\!\theta_{W}^{}c_{\tilde{W}W}'\big] .
\end{align}
\label{eq:on-shell_coefficients} 
\end{subequations}
\hspace*{-1.5mm}Corresponding to the number of coefficients, only 4 independent operators
contribute: $\O_{\tilde{B}W}^{\prime}$, $\O_{\tilde{B}B}^{\prime}$,
$\O_{\tilde{W}W}^{\prime}$, and $\O_{\tilde{B}W}^{}\!-\!\O_{\tilde{W}B}^{}\hs$.\ 
One notable feature is that the contributions of $\O_{\tilde{W}W}$ and $\O_{\tilde{B}B}$
are negligible in the nTGC on-shell production of gauge bosons,
since they only contribute to vertex terms that are proportional to $q^{\mu}_V$, and 
result in negligible amplitudes suppressed by the incoming fermion mass, 
as discussed in the text below Eq.\eqref{eq:on-shell_vertices}.\ 
{It turns out that these 4 operators $\O_{\tilde{B}W}^{\prime}$, $\O_{\tilde{B}B}^{\prime}$,
$\O_{\tilde{W}W}^{\prime}$, and $\O_{\tilde{B}W}^{}\!-\!\O_{\tilde{W}B}^{}\hs$ 
are also the only operators that contribute to the off-shell $V^*V^*$ production at colliders, 
where each gauge boson $V$ decays into two fermions subsequently.\ 
This is further discussed in Appendix\,\ref{apd:off-shell_vertices}.}

\vspace*{0.7mm}

When the loop diagrams of the UV theory that generate nTGCs contain
only heavy particles with masses of the order of the cutoff scale $\Lambda$, 
the operators \eqref{eq:operators1}-\eqref{eq:operators2} and
thus Eq.\eqref{eq:on-shell_coefficients} include the complete contributions
of order $1/\Lambda^{4}$.\  However, when a loop diagram contains both heavy
and light particles,  it contains soft parts that are not contained
in Eqs.\eqref{eq:operators1}-\eqref{eq:operators2}.\ 
The soft parts of the loop diagrams contain logarithmic dependences 
on the external momentum that violate the conditions enforcing the form of Eq.\eqref{eq:on-shell_vertices}.\
For the SMEFT in the IR region, it must be accommodated by a loop diagram by contracting light fermionic fields in
effective operators obtained by tree-level matching.\
In this case, the full one-loop contribution of the UV theory is captured by two types of contributions in the SMEFT: 
1)\,the tree-level contribution from the nTGC operators \eqref{eq:operators1}-\eqref{eq:operators2} 
obtained by one-loop matching, and 2)\,the one-loop diagram of operators involving fermionic fields 
obtained from tree-level matching.\ 
This point will be discussed in detail in the following sections.
\begin{figure}[]
\centering
\includegraphics[width=0.84\textwidth]{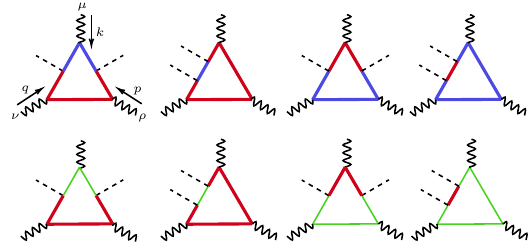}
\vspace*{1mm}
\caption{\small 
Two classes of one-loop fermionic contributions to the nTGCs, including the pure heavy fermion loops
as shown in the first row, or including the mixed heavy-light fermion loops as shown in the second row.\  
In each class there are four types of fermionic loop diagrams that contribute to the nTGCs.\  
The dashed lines denote Higgs fields, the thin solid lines denote $E$ propagators, 
the thick solid lines denote $N$ propagators, and the wiggly lines denote gauge bosons.\ 
The sum over directions of fermion flow is implied for each diagram.} 
\label{fig:basic_diagrams}
\label{fig:1}
\end{figure}

\section{\large\hspace*{-2.5mm}Structure of Heavy Fermion Loop Contributions to nTGCs}
\label{sec:loop_structure}
\label{sec:3}

\begin{figure}[t]
\vspace*{-5mm}
\centering
\includegraphics[width=0.90\textwidth]{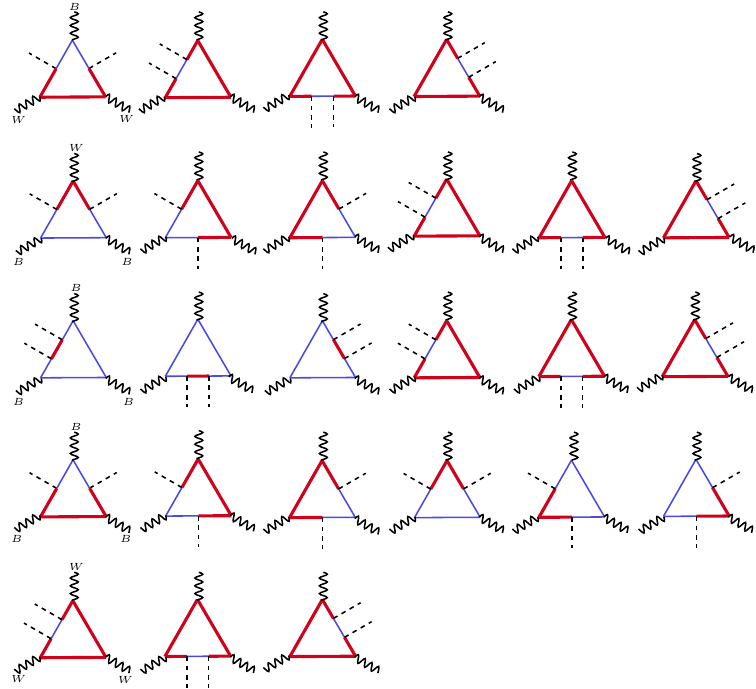}
\caption{\small 
List of the relevant one-loop contributions to the four types of nTGC vertices,
including the vertices of $BWW$ (1st row), $WBB$ (2nd row), $BBB$ (3rd and 4th rows), 
and $WWW$ (5th row).\  The loop structures are those shown in Fig.\,\ref{fig:1}, 
where for each diagram a sum over directions of the fermion loop flows is implied.\
In each diagram, the dashed lines denote Higgs fields, the thin solid lines denote $E$ propagators, 
the thick solid lines denote $N$ propagators, and the wiggly lines denote gauge bosons.\ 
} 
\label{fig:2}
\vspace*{-2mm}
\end{figure}
\begin{figure}[h]
\centering
\includegraphics[width=0.25\textwidth]{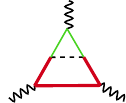}
\caption{\small 
A sample two-loop diagram containing internal fields of heavy fermions and Higgs doublet
that contribute to the nTGCs.\  
The dashed lines denote Higgs fields, the thin solid lines denote $E$ propagators, 
the thick solid lines denote $N$ propagators, and the wiggly lines denote gauge bosons.\ 
Here a sum over directions of the fermion loop flows is implied.} 
\label{fig:3}
\end{figure}

Since the Lorentz structures of the relevant dimension-8 operators incorporate the Levi-Civita tensor, 
we restrict ourselves to extensions of the SM with heavy fermions.\  
We consider the Yukawa interaction between 
a fermionic weak doublet $N$ and a fermionic weak singlet $E$ 
with hypercharges $Y_{N}^{}$ and $Y_{E}^{}\!=\!Y_{N}^{}\!-\!1/2$, respectively.\ 
The interaction takes the following form:
\begin{equation}
\bar{N} H(c_{V}^{} \!+\hsm c_{A}^{}\gamma_{5}^{})E + \rm{h.c.}
\label{eq:general_model} \
\end{equation}
The mass scales of the fields $N$ and $E$ differs in the two
scenarios that we are going to consider.\
(i)\,In the ``all-heavy'' case
both $N$ and $E$ are heavy vector-like particles to be integrated out at low energies.\ 
For simplicity of calculation, we choose the two
fermions to have similar masses $m_{N}^{}\simeq m_{E}^{}\simeq\! M$, and
the mass difference plays a negligible role.\  In this way, we need to 
only deal with an EFT having a single heavy mass scale $M$.\ 
(ii)\,In the ``heavy-light''
case only one of $N$ or $E$ is the heavy vector-like fermion to be integrated out 
at a heavy mass scale $M$, whereas the other one is a light chiral fermion
in the SM.\  In this scenario, we set $c_{A}^{}\!=\!\pm c_{V}^{}$  to project
out the chiral component.\  Both cases can be realized by well motivated new physics models.\ 
For instance, the Higgsino-Bino system in the Minimal Supersymmetric SM (MSSM) corresponds to
the ``all-heavy'' case, whereas the models with a heavy right-handed
neutrino correspond to the ``heavy-light'' case.\  We leave the
details of the model discussions to Section\,\ref{sec:results}, after elaborating the
methodology for the EFT matching in Section\,\ref{sec:method}.

There are four types of topology for loop diagrams that contributes to nTGCs, as
illustrated in Fig.\,\ref{fig:basic_diagrams}.\  We denote  Higgs fields by dashed lines,
the $E$ field by a thin solid line, the $N$ field by a thick solid line, and the gauge bosons  
by wiggly lines.\
The momenta of the external Higgs fields will be set to zero 
for our purpose of deriving the nTGC vertices.\footnote{
We perform the matching by computing diagrams for $H_0H_0VVV$\,(with $H_0$ denoting the Higgs VEV and carrying zero external momentum) because they unambiguously correspond to nTGC vertices.\ It is also possible to perform the matching with $HHVV$ box diagrams where $H$ carries non-zero momentum in the symmetric phase, following the method of 
\cite{Ren:2022tvi}.\ However, this approach may contain contributions that are irrelevant to the dimension-8 nTGC operators and need to be distinguished carefully.\ 
This is worthy of future investigation.}\
The sum over directions of fermion flow is implied in each diagram.\ 
The results for different models are obtained by including the associated tensor 
structures and couplings in the model.\ 
The relevant diagrams for each type of vertex in the gauge 
eigenbasis are listed in Fig.\,\ref{fig:2}.\ The coefficient of each 
nTGC operator is then determined by matching the result from 
the UV theory with these loop diagrams and vertices.

We showed in Section\,\ref{sec:general_operators} that Higgsless (pure
gauge) nTGC operators cannot be obtained from the fermionic one-loop diagrams in
a renormalizable model.\ However, they can be generated at the two-loop level,
e.g., by contracting the two Higgs fields attached to the fermion loop
in the diagrams of Fig.\,\ref{fig:1}, and these two-loop contributions are allowed 
by the known symmetries of the UV model.\ A sample diagram of such two-loop contribution
is shown in Fig.\,\ref{fig:3}.\  
Although a full discussion of the two-loop nTGC vertex lies
beyond the scope of this paper, we note that it may be obtained by contracting
the two Higgs fields of the one-loop effective operators \eqref{eq:operators1}
and \eqref{eq:operators2}.\ 
%

\section{\large\hspace*{-2.5mm}Matching to UV Completion and Induced nTGCs}
\label{sec:method}
\label{sec:4}

Since we are interested in momenta much smaller than the heavy fermion mass $M$,
the loop integral can be approximated with the method of regions, which
is really convenient for the matching calculation of EFT coefficients\,\cite{Chetyrkin:1988zz}-\cite{Semenova:2018cwy}\cite{Georgi:1993mps}.\footnote{%
A short introduction of this method in the context of EFT was given in Ref.\,\cite{Cohen:2019wxr}.} 
In the following, we briefly review the application of this
method to the two scenarios that we will consider.

In the following, we denote the loop diagrams to be evaluated by $\Gamma_{i}$.\ 
If $\Gamma_{i}$ contains only large mass propagators as in the ``all-heavy'' case, it is sufficient to expand directly the integrand, treating
external momenta as small variables.\ Since the loop momentum $\ell$
mainly contributes when $\ell\!\sim\! M$ in this case, the expansion is
finite at all orders in the external momenta.\ However, when the integrand of
$\Gamma_{i}$ contains both massive and massless propagators as in
the ``heavy-light'' case, the loop integrals receive contributions
from both the hard region with $\ell\!\sim\!M$ and the soft region with $\ell\!\ll\!M$.\ 
In general, the method of regions splits the integral $\Gamma_{i}$ into
a soft part and a hard part:
\begin{equation}
\Gamma_{i}\,=\,\Gamma_{i}\bigg|_{\text{hard}}+\Gamma_{i}\bigg|_{\text{soft}}\text{\,.}
\end{equation}
The hard piece is obtained by expanding the integrand in $\Gamma_{i}$
by taking the external momenta $(k,p,q)$ as small variables (assuming
all other masses are negligibly small), and by treating the loop momentum
$\ell$ and the mass $M$ as large quantities.\ 
Thus, a propagator in the hard piece may be expanded as 
\begin{subequations}
\begin{align}
\frac{\text{i}}{\,(\ell+p')^{2}\!-\!M^{2}\,}\bigg|_{\text{hard}} & =\, 
\frac{\text{i}}{~\ell^{2}\!-\!M^{2}~}-\frac{\text{i}(2\ell\hsm\cdot\hsm p')}
{~(\ell^{2}\!-\!M^{2})^{2}~}+\cdots\,,
\\[1mm]
\frac{\text{i}}{~(\ell+p')^{2}~}\bigg|_{\text{hard}} & =\,
\frac{~\text{i}}{\ell^{2}~}-\frac{~\text{i}(2\ell\hsm\cdot\hsm p')~}
{(\ell^{2})^{2}}+\cdots\,,
\end{align}
\end{subequations}
 where $p'$ is a linear combination of external momenta that enters
the propagator.\ The soft part is obtained by treating the loop momentum
as a small expansion variable like the external momenta, so a massive
propagator expands as follows for the soft part:
\begin{equation}
\frac{\text{i}}{~(\ell+p')^{2}\!-\!M^{2}~}\bigg|_{\text{soft}}
=-\frac{\text{i}}{\,M^{2}\,}-\frac{~\text{i}(\ell+p')^{2}~}{M^{4}}+\cdots\,.
\end{equation}
In other words, the massive propagator shrinks to a point in the soft region, 
just as in the EFT obtained by tree-level matching.\ On the other hand, the
massless propagators stay unchanged in the soft piece. 

The hard part and
the soft part mostly capture the contributions from $\ell\!\sim\! M$ and
$\ell\!\sim\! p,k,q$, respectively.\ But, they also modify the behaviors
of the integrand in the IR and in the UV.\ Compared to the original integral,
the hard piece raises the power of $\ell$ in the denominator of the
integrand and may render the integral divergent as $\ell\!\rightarrow\!0\hs$.\ 
Similarly, the soft piece may contain a divergence as $\,\ell\!\rightarrow\!\infty\hs$.\ 
The key point of the method of regions and the EFT calculation with
dimensional regularization is that these modifications cancel each
other, so that the two artificial divergences introduced by the expansion
cancel precisely between the soft and hard pieces\,\cite{Georgi:1993mps}.\
Hence, the total result of evaluating the loop diagram is
finite, as expected from power counting.

This method is really useful for the EFT matching.\ The soft part of the
loop diagram is equivalent to a loop diagram that closes lines of
light fields in the tree-level EFT obtained by shrinking heavy propagators to points 
in the full theory.\  The hard part then only contributes
to the one-loop effective operator that compensates the difference
between the evaluation in the tree-level EFT (soft part) and the full
one-loop evaluation.\ For a brief review of this process, we denote the
full UV theory as $\,{\cal L}^{U}\!={\cal L}_{0}^{U}+{\cal L}_{\text{ct}}^{U}\,$
and the effective theory after integrating out the heavy fields as
${\cal L}_\text{EFT}=\bar{{\cal L}}^{(0)}+\bar{{\cal L}}^{(1)}$, where
$\,\bar{{\cal L}}^{(i)}\!=\bar{{\cal L}}_{0}^{(i)}\!+\bar{{\cal L}}_{\text{ct}}^{(i)}$
is the EFT terms from $i\hs$th-loop matching, including the bare terms
and the corresponding counter terms.\ Then, the one-loop matching condition
for a process $P$ of light fields is
\begin{equation}
\Gamma_{P}^{(1)}({\cal L}^{U})\,=\,
\Gamma_{P}^{(1)}(\bar{{\cal L}}^{(0)})+\Gamma_{P}^{(0)}(\bar{{\cal L}}^{(1)})\hs,
\end{equation}
where $\Gamma_{P}^{(i)}({\cal L})$ is the sum of 1-light-particle
irreducible diagrams of the (off-shell) process $P$ at the $i\hs$th-loop
order from the interaction terms of ${\cal L}\hs$.\ The tree-level matching
procedure determines the tree-level EFT Lagrangian $\bar{{\cal L}}^{(0)}$
and ensures $\Gamma_{P}^{(1)}({\cal L}^{U})\big|_{\text{soft}}\!
\!=\!\Gamma_{P}^{(1)}(\bar{{\cal L}}^{(0)})\hs$.\ 
The remaining hard part matches
\\[-5mm]
\begin{equation}
\Gamma_{P}^{(1)}({\cal L}^{U})\bigg|_{\text{hard}}
=\, \Gamma_{P}^{(0)}(\bar{{\cal L}}^{(1)})\,.\label{eq:hard_part_matching}
\end{equation}
This determines the one-loop terms of the EFT Lagrangian.


In the case of the nTGC
loop diagram of Fig.\,\ref{fig:1} in a heavy
fermion model ${\cal L}^{U}$, the one-loop EFT operators $\bar{{\cal L}}^{(1)}$
are just those of Eqs.\eqref{eq:operators1}-\eqref{eq:operators2}
and their corresponding counter terms, with Wilson coefficients to
be determined by the matching procedure.\ Although $\Gamma_\text{nTGC}^{(1)}({\cal L}^{U})$
is finite, the intermediate variables $\Gamma_\text{nTGC}^{(1)}({\cal L}^{U})\big|_{\text{soft}}$
and $\Gamma_\text{nTGC}^{(1)}({\cal L}^{U})\big|_{\text{hard}}$ contain
artificial divergences because of the propagator expansions.\ In the
matching procedure, the divergence in $\Gamma_\text{nTGC}^{(1)}({\cal L}^{U})\big|_{\text{soft}}$
corresponds to the divergence in the EFT diagram $\Gamma_\text{nTGC}^{(1)}(\bar{{\cal L}}^{(0)})$,
and the divergence in $\Gamma_{P}^{(1)}({\cal L}^{U})\big|_{\text{hard}}$
corresponds to the counter term vertex $\Gamma_\text{nTGC}^{(0)}(\bar{{\cal L}}_{\text{ct}}^{(1)})\hs$.\
The divergences cancel in both the EFT evaluation and the full theory
evaluation. 


The artificial divergences introduced by the method of regions in the
intermediate steps require caution in the treatment of $\hs\gamma_{5}^{}\hs$,
since it does not have a natural definition in dimensional regularization
with $D\neq4$. This is not an issue for the ``all-heavy'' case, 
but needs care for the ``heavy-light'' case, since the latter
contains intermediate divergences that cancel between the hard and
soft parts.\ Several schemes for treating $\hs\gamma_{5}^{}\hs$ have been
developed. One needs to sacrifice either the anti-commutativity of
$\hs\gamma_{5}^{}\hs$ with all the other $\gamma$ matrices as in the Breitenlohner-Maison-'t
Hooft-Veltman (BMHV) scheme~\cite{tHooft:1972tcz}-\cite{Belusca-Maito:2020ala}, 
or give up the cyclic property of the trace of a string of $\gamma$-matrices
with an odd number of $\gamma_{5}^{}$ matrices by treating the trace as a projection
operation, as in naive dimensional regularization 
(NDR)\,\cite{Kreimer:1989ke}-\cite{Kreimer:1993bh}.\footnote{See Ref.\,\cite{Jegerlehner:2000dz} 
for an overview on the $\gamma_{5}^{}$ issue in dimensional regularization.}\ 
The BMHV scheme was shown to be self-consistent to all perturbative orders\,\cite{Breitenlohner:1975hg}-\cite{Breitenlohner:1977hr}.\
However, the non-commutativity between $\gamma_{5}^{}$ and some of the
$\gamma_{\mu}^{}$ gives rise to intermediate gauge symmetry-violating terms 
and needs counter terms in the renormalization procedure to restore gauge independence 
of the physical result.\ In the context of EFT, the intermediate
gauge symmetry violation in the BMHV scheme is manifested in irrelevant anomalies 
that are removable by finite counter terms\,\cite{Feruglio:2020kfq}-\cite{Cohen:2023gap}.


In an attempt to maintain gauge invariance in the intermediate step and
for the convenience of calculation, we adopt the NDR scheme that
defines an anticommuting $\gamma_{5}^{}$ as follows: 
\begin{equation}
\{\gamma_{5},\gamma_{\mu}\}=0 \hs, 
\end{equation}
for all $\mu\!\leqq\! D\hs$. This scheme was shown to maintain gauge invariance
automatically without the need of further counter terms
at least in one-loop order\,\cite{Kreimer:1993bh}.\  To be clear,
we explain concisely the practical procedure of NDR\,\cite{Kreimer:1993bh} 
as applied to our calculation.\   One cannot simply continue the relation 
$\text{tr}(\gamma_{\mu_{1}}^{}\!\cdots\gamma_{\mu_{4}}^{}\gamma_{5}^{})
=\text{i}4\epsilon_{\mu_{1}\cdots\mu_{4}}$
to $D\!\neq\!4$ in DREG, since the rank-4 anti-symmetric tensor
is defined in $4$-dimensional spacetime only.\ The NDR scheme
treats the tensor $\epsilon_{\mu_{1}^{}\cdots\mu_{4}^{}}$ in $D\!\neq\!4$ 
as a regular rank-4 tensor rather than an anti-symmetric tensor.\
The trace containing an odd number of $\gamma_{5}$ for $D\!\neq\!4$, 
$\text{tr}(\gamma_{\mu_{1}}^{}\cdots\gamma_{\mu_{2n}}^{}\gamma_{5}^{})$,
is regarded as a projection operation that happens to
give the same result as $D\!=\!4$, and is not a trace of matrices anymore.\ 
Hence, it loses its cyclic property, and requires a consistent
choice of ``reading point'' to write down the order of the $\gamma$ matrices in the chain.\  
Hereafter, the ``reading point'' refers
to the last matrix that appears in the trace whenever there is odd
number of $\gamma_{5}^{}$.\ In this work, we choose one of the Higgs
vertices as the reading point in all the loop calculations.\ 
The four types of diagrams in Fig.\,\ref{fig:basic_diagrams} then become:
%
\begin{align}
\Gamma_{(a)}^{} & =\!\int\!\!\!\frac{d^{4}\ell}{(2\pi)^{4}}\text{tr}\!
\left[G_{N}(\ell)\gamma^{\rho}G_{N}(\ell\!-\!p)\gamma^{\nu}G_{N}(\ell\!+\!k)V_{H}^{+}G_{E}(\ell\!+\!k)
\gamma^{\mu}G_{E}(\ell)V_{H}^{-}\right]\!\! +\!(\text{reverse\,fermion\,flow}),
\nn\\
\Gamma_{(b)} & =\!\int\!\!\!\frac{d^{4}\ell}{(2\pi)^{4}}\text{tr}\!
\left[G_{N}^{}(\ell)\gamma^{\mu}G_{N}(\ell\!-\!k)\gamma^{\rho}G_{N}^{}(\ell\!+\!q)
\gamma^{\nu}G_{N}^{}(\ell)V_{H}^{+}G_{E}(\ell)V_{H}^{-}\right]\!\!+\!(\text{reverse\,fermion\,flow}),
\nn\\
\Gamma_{(c)} & =\!\int\!\!\!\frac{d^{4}\ell}{(2\pi)^{4}}\text{tr}\!
\left[G_{N}(\ell\!+\!k)\gamma^{\mu}G_{N}(\ell)V_{H}^{+}G_{E}(\ell)\gamma^{\rho}G_{E}(\ell\!-\!p)
\gamma^{\nu}G_{E}^{}(\ell\!+\!k)V_{H}^{-}\right]\!\!+\!(\text{reverse\,fermion\,flow}),
\nn\\
\Gamma_{(d)} & =\int\frac{d^{4}\ell}{(2\pi)^{4}}\text{tr}\left[G_{N}(\ell)V_{H}^{+}G_{E}(\ell)\gamma^{\mu}G_{E}(\ell-k)\gamma^{\rho}G_{E}(\ell+q)\gamma^{\nu}G_{E}(\ell)V_{H}^{-}\right]+(\text{reverse fermion flow}),
\label{eq:base_loop}
\end{align}
\\[-5.4mm]
\hspace*{-2mm}
where $V_{H}^{\pm}\!=\!(c_{V}^{}\!\pm\! c_{A}^{}\gamma_{5}^{})$ are the vertices that
connect to the Higgs fields and $G_{N,E}(p)$ are the propagators of the heavy fermion fields 
$N$ and $E$ with momentum $p\hs$, respectively.\ In this expression, $V_{H}^{-}$ is the
reading point of the trace and this choice will persist through all
fermionic loop evaluations, including the ones abbreviated by ``reverse fermion flow''.\ 
Once the traces are written by following the same reading
points ($V_{H}^{-}$), the projection operation is performed by {first}
moving $\gamma_{5}$ to the end of the trace using its anti-commutative
property and {then} making the replacement: 
\begin{equation}
\gamma_{5}^{}\,\rightarrow\,
-\frac{\text{i}}{24}\epsilon_{\mu\nu\rho\sigma}\gamma^\mu\gamma^\nu\gamma^\rho\gamma^\sigma\,.
\label{eq:NDR_G5_replacement}
\end{equation}
Unlike in 4 dimensions, in NDR this procedure should not be regarded as a definition
of $\gamma_{5}^{}\hs$, but rather as a handy way to compute the result
of the projection denoted as $\text{tr}(\cdots\gamma_{5})$ \textit{after}
following a strict reading point prescription and anti-commuting $\gamma_{5}$
to the end of the trace\,\cite{Kreimer:1993bh}.\ 
The tensor $\epsilon_{\mu\nu\rho\sigma}^{}$ becomes fully antisymmetric only when taking the limit 
$D\!\to\!4\,$.

\vspace*{0.8mm}

The non-cyclicity of $\text{tr}(\cdots\gamma_{5})$ is proportional to $\hs\epsilon \!=\! (4\!-\!D)/2$
and vanishes under $D\!\to\!4\hs$.\ 
Thus, it manifests itself in the limit $D\!\rightarrow\!4$ only by cancelling the
$\hs 1/\epsilon\,$ pole in the divergent term.\   But, all the diagrams in both
the ``all-heavy'' and ``heavy-light'' cases are finite, so the
non-cyclicity will not play a role in the final physical nTGC vertex
function, as long as the reading point is kept consistent between the
soft and hard parts of the same diagram so that their intermediate
divergences cancel precisely.\  The situation becomes more subtle
when matching the diagrams to an EFT in the ``heavy-light case'',
where the one-loop effective operators and their counter terms match
to the divergent hard part of the diagram as described in Eq.\eqref{eq:hard_part_matching}.
For some choices of reading points, the non-cyclicity of
the trace combines with the divergence and appears as a finite term
in the result.\ This term may break the manifest gauge invariance of
the hard and soft parts separately (but not their sum), so that
the finite terms of the hard part do not match to a set of gauge-invariant
operators.\ This would be the case if we had chosen the gauge vertices
as the reading points, in which case the matching procedure might require an additional
set of finite gauge-violating counter terms.\  This would impair the convenience
of choosing NDR over BMHV.\ Fortunately, as we show below, choosing the Higgs
vertices as reading points is free of these technical issues.\ 
In these cases the hard part can be matched directly to a set of gauge-invariant operators 
and the soft part alone satisfies the Ward-Takahashi identity.\ 
This particular choice of reading point combined with the
NDR treatment of $\gamma_{5}^{}$ is a renormalization scheme that preserves
manifest gauge invariance in all the intermediate steps of the EFT
matching problem.\

\begin{figure}[t] 
\begin{center} 
\includegraphics[width=0.8\textwidth]{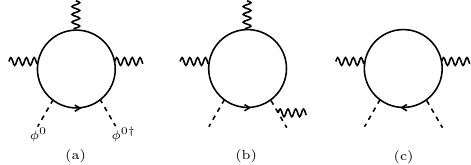}
\vspace*{-2mm}
\caption{\small Sample diagrams that enter the Ward-Takahashi identity
\eqref{eq:WT_position_space}.\ Plot (a) has 3 external gauge bosons attached to a fermion loop.\ 
Plot (b) has 2 vector bosons attached to the fermion loop and one to the Higgs line.\ 
Plot (c) gives the correlation function involving 2 gauge bosons that enter the 
left-hand side of Eq.\eqref{eq:WT_position_space}.
In each diagram, the dashed lines denote Higgs fields, the solid lines denote fermion propagators, 
and the wiggly lines denote gauge bosons.\ 
} 
\label{fig:WT_fig}
\label{fig:4}
\end{center}
\vspace*{-5mm}
\end{figure}

In the following, we consider the Ward-Takahashi identity for $U(1)_{B}^{}$
and $U(1)_{T_3^{}}^{}$ related to the loop diagrams of Fig.\,\ref{fig:basic_diagrams}
to illustrate the necessity of choosing Higgs vertices as the reading points.\ 
For a loop diagram that generates the $V^{\mu}$-$V_{1}^{\nu}$-$V_{2}^{\rho}$ vertex, 
gauge invariance enforces the following identity for the position space correlators:
\begin{align}
\label{eq:WT_position_space}
&\frac{\partial}{\partial x^{\mu}}\langle J_{V}^{\mu}\hsm (x)
J_{\!f_{1}}^{\nu}\hsm (y)J_{\!f_{2}}^{\rho}(z)
\phi^{0\hs\dagger}(u)\phi^{0}(v)\rangle_{T}^{} 
\\[0.6mm] 
& =\!\langle J_{\!f_{1}}^{\nu}\hsm (y)J_{\!f_{2}}^{\rho}\hsm (z)
\hsm\big[Q_{V}^{\phi^{0}}\delta^{(4)}\hsm (x\!-\!u)\big]
\hsm\phi^{0\hs\dagger}(u)\phi^{0}(v)\rangle_{T}^{} 
\!+\!\langle J_{\!f_{1}}^{\nu}\hsm (y)J_{\!f_{2}}^{\rho}\hsm (z)\hsm 
\big[\hsm\!-\hsmx Q_{V}^{\phi^{0}}\delta^{(4)}\hsm (x\!-\!v)\big]\hsm 
\phi^{0\hs\dagger}(u)\phi^{0}(v)\rangle_{T}^{}\, .
\nn 
\end{align}
where $\langle\hsm\cdots\rangle_{T}$ is the time-ordered vacuum expectation value 
and  $\phi^{0}$ denotes the neutral component of the SM Higgs doublet field $H\hs$.\  
In the above, 
$J_{V}^{\mu}$ ($V\!\!=\!B,W^{0}\hs$) denotes the conserved current of $U(1)_{B}^{}$ and $U(1)_{T_{3}}^{}$, 
and $J_{f_{i}}^{\nu}\!=\!\bar{f_{i}}\gamma^{\nu}\hsm f_{i}^{}$ is the current coupled to 
the external gauge boson $V_i^\nu$,
where $f_{i}^{}$ denotes the fermions in the loop.\  The correlator on the left-hand
side of Eq.\eqref{eq:WT_position_space} corresponds to diagrams with
three gauge bosons, including the case (a) with all three gauge bosons attached
to the fermion loop, as in Fig.\,\ref{fig:WT_fig}(a); and case (b) with 
two gauge bosons $V_{1}^{\nu}$ and $V_{2}^{\rho}$ attached to the
loop at vertices described by $J_{\nu}^{f_{1}}$ and $J_{\rho}^{f_{2}}$,
and the third gauge boson $V^{\mu}$ attached to a Higgs line via the Higgs current part of $J_{V}^{\mu}$, 
as in Fig.\,\ref{fig:WT_fig}(b).\ 
The correlators on the right-hand side correspond to loops with two
gauge vertices (represented by $J^{f_{1}}$ and $J^{f_{2}}$) on the fermion loop, 
as in Fig.\,\ref{fig:WT_fig}(c).\ 
Fourier-transforming the identity to momentum space, subtracting the Fig.\ref{fig:WT_fig}(b)-type
diagrams with the gauge boson $V^{\mu}$ attaching to $\phi_{0}^{}\hs$, and amputating the
$\phi_{0}^{}$ propagator, we derive the following Ward-Takahashi identity 
for the amputated amplitudes:
\\[-7.5mm]
\begin{align}
& k_{\mu}^{}\AA_{V}^{\mu\nu\rho}(k;p_{1}^{},p_{2}^{};p_{\phi}^{}\!=\!0,p_{\phi^{\dagger}}\!=\!0) 
\nn\\[-3mm]
\label{eq:WT_momentum_space}
\\[-3mm]
& =-Q_{V}^{\phi_{0}}\big[\AA_{0}^{\nu\rho}(p_{1},p_{2};p_{\phi}^{}\!=\!0,p_{\phi^{\dagger}}\!=\!k)
-\AA_{0}^{\nu\rho}(p_{1}^{},p_{2}^{};p_{\phi}^{}\!=\!k,p_{\phi^{\dagger}}\!=\!0)\big],
\nn 
\end{align}
where $(k,\hs p_{1}^{},\hs p_{2}^{},\hs p_{\phi}^{},\hs p_{\phi^{\dagger}}^{})$
are the momenta obtained by Fourier-transforming the position variables 
$(x, y, z, u, v)$ in Eq.\eqref{eq:WT_position_space},
all defined with directions going into the loop, and $p_{\phi}$ and
$p_{\phi^{\dagger}}$ are momenta going into the loop via the $\phi_{0}^{}$
and $\phi_{0}^{\dagger}$ lines.\ 
The left-hand side of Eq.\eqref{eq:WT_momentum_space}
includes only the diagrams of the Fig.\ref{fig:WT_fig}(a)-type with three gauge
vertices attached to the fermion loop (rather than the Higgs line), 
exactly like the diagrams of Fig.\,\ref{fig:1}. The correlators on the
right-hand side of Eq.\eqref{eq:WT_momentum_space} are of the type of Fig.\,\ref{fig:WT_fig}(c).

The hard part of a set of nTGC loop diagrams can be matched to a set
of gauge-invariant operators only when their corresponding Ward-Takahashi
identity \eqref{eq:WT_momentum_space} still holds with the
involved amplitudes in the identity restricted to their hard parts:
\begin{equation}
\AA ~\to~ \AA_{\text{hard}}^{}\,.
\end{equation}
If this is the case, the hard-part version of Eq.\eqref{eq:WT_momentum_space}
would correctly connect the coefficients of nTGC couplings to 
the 2-gauge-boson vertex couplings induced by the same set of operators \eqref{eq:operators1}
and \eqref{eq:operators2}.\   But, as mentioned above, the choice of
the reading point of a trace that contains an odd number of $\gamma_{5}^{}$
could break the hard-part version of \eqref{eq:WT_momentum_space}
if the identity involves cancellations between cyclic permutations
of matrices in the trace together with a divergent hard-part integral.\ 
In the following, we show that choosing the Higgs vertices as reading
points ensures that the identity \eqref{eq:WT_momentum_space} holds without the
need of trace cyclicity.

\vspace*{0.8mm}

We can write the left-hand side of the amplitude of Eq.\eqref{eq:WT_momentum_space}
with a Higgs reading point (taken as $\phi^{0}$ for example)
as follows:
\begin{align}
& k_{\mu}\AA_{V}^{\mu\cdots}(k;\{p_{\{i\}}\};p_{\phi}^{}\!=\!0,p_{\phi^{\dagger}}^{}\!=\!0)
\nn\\
&\hspace*{10mm}
=\sum_{\{p_{\{i\}}^{a}\}}\!\int\!\!\!\frac{d^{4}q}{\left(2\pi\right)^{4}}
\text{tr}[\M_{b}(q';\{p_{\{i\}}^{b}\})V_{\phi^{\dagger}}k_{\mu}
\bar{\M}_{a}^{\mu}(q;k,\{p_{\{i\}}^{a}\})V_{\phi}]
\nn\\
&
\hspace*{15mm} +\sum_{\{p_{\{i\}}^{a}\}}\!\int\!\!\!\frac{d^{4}q}{\left(2\pi\right)^{4}}
\text{tr}\!\left[k_{\mu}\bar{\M}_{b}^{\mu}(q';k,\{p_{\{i\}}^{b}\})V_{\phi^{\dagger}}
\M_{a}(q;\{p_{\{i\}}^{a}\})V_{\phi}\right]\!.
\label{eq:WT_check1}
\end{align}
Here we generalize triple gauge-boson amplitudes to amplitudes with any number
of external gauge bosons, and suppress all the gauge indices except the one
to be contracted with $k^{\mu}$,  and $V_\phi^{}$ and $V_{\phi^{\dagger}}$
are the two Yukawa vertices that connect to the external Higgs lines, 
while $\{p_{\{i\}}\}$ represents the set of gauge boson momenta (other than $k^{\mu}$).\   
We have separated the loop into two blocks sandwiched by the two Yukawa vertices, 
denoted by $\M_{a}$ and $\M_{b}$, or $\bar{\M}_{a}^{\mu}$ and $\bar{\M}_{b}^{\mu}$ 
(with a $V^{\mu}(k)$ inserted into a propagator therein).\ 
The sum of $\{p_{\{i\}}^{a}\}$ runs over all possible gauge boson momenta except $k^{\mu}$, 
which enters $\M_{a}$ or $\bar{\M}_{a}^{\mu}$.\   The first argument ($q$ or $q'$) in $\M$ or
$\bar{\M}^{\mu}$ represents the momentum of the first fermion propagator
in the block that leaves the Higgs vertex and appears in the block.\ 
For the first line of Eq.\eqref{eq:WT_check1} we have $q'\!=\! q\!+\!\sum_{i}p_{i}^{a}\!+\!k\hs$
and for the second line we have $q'\!=\!q\!+\!\sum_{i}p_{i}^{a}\hs$.\ 
Since we are considering U(1) currents and the fields are in their gauge eigenstates, the
fermion species remains unchanged in each block, and we denote them
as $f_{a}$ and $f_{b}$ in block $a$ and $b$ respectively.\ 
Another Ward-Takahashi identity similar to that of QED gives 
\begin{subequations}
\begin{align}
k_{\mu}\bar{\M}_{a}^{\mu}(q;k,\{p_{\{i\}}^{a}\}) 
& \,=\, 
-Q_{V}^{f_{a}}\!\left[\M_{a}(q;\{p_{\{i\}}^{a}\})\!-\!\M_{a}(q\!+\!k;\{p_{\{i\}}^{a}\})\right]
\!,
\\[0.8mm]
k_{\mu}\bar{\M}_{b}^{\mu}(q';k,\{p_{\{i\}}^{b}\}) 
& \,=\,
-Q_{V}^{f_{b}}\!\left[\M_{b}(q';\{p_{\{i\}}^{b}\})\!-\!\M_{b}(q'\!+\!k;\{p_{\{i\}}^{b}\})\right]\!.
\end{align}
\label{eq:WT_QED} 
\end{subequations}
\hspace*{-2mm}
Substituting these identities into Eq.\eqref{eq:WT_check1} and using 
the relation $Q_{V}^{f_{a}}\hsm +\hsm Q_{V}^{\phi^{0}}\!=\!Q_{V}^{f_{b}}$, 
we deduce the following identity:
\begin{align}
& k_{\mu}\AA_{V}^{\mu\cdots}(k;\{p_{\{i\}}\};p_{\phi}\!=\!0,p_{\phi^{\dagger}}\!=\!0) 
\nn\\[-3.1mm]
\\[-3.1mm]
&=
Q_{V}^{\phi_{0}}\!\left[\AA_{0}^{\cdots}(\{p_{\{i\}}\};p_{\phi}\!=\!k,p_{\phi^{\dagger}}\!=\!0)
\!-\AA_{0}^{\cdots}(\{p_{\{i\}}\};p_{\phi}\!=\!0,p_{\phi^{\dagger}}\!=\!k)\right]\!.
\nn 
\end{align}
For the case of three gauge bosons, this reduces to the identity \eqref{eq:WT_momentum_space}.\ 
We see that for the identity \eqref{eq:WT_momentum_space} to hold, it is sufficient to validate
the QED-like Ward-Takahashi identities \eqref{eq:WT_QED}  
for each block sandwiched between the Yukawa vertices $V_{\phi}^{}$
and $V_{\phi^{\dagger}}^{}$.\   When choosing Yukawa matrices as the reading
points, these blocks are not wrapped around the ends of traces, and thus
one does not need cyclicity to prove Eq.\eqref{eq:WT_momentum_space}.\ 
It is also apparent why choosing another vertex as reading point may
violate the hard-part version of Eq.\eqref{eq:WT_momentum_space}.\ 
The diagrammatic proof
of Eq.\eqref{eq:WT_QED}\footnote{See for example a textbook derivation of this  
in Chapter\,7 of Ref.\,\cite{Peskin:1995ev}.} sums over all possible insertions of $V^{\mu}(k)$
into all propagators involved.\ 
For instance, in the right-hand side of Eq.\eqref{eq:WT_check1}, 
if another vertex within $\bar{\M}_{a}^{\mu}$ to the right of the 
$V^{\mu}(k)$ insertion was chosen as the reading point, 
the first trace on the right-hand side of Eq.\eqref{eq:WT_check1} would take the following form,  
\begin{equation}
\text{tr}\hsm\big[\bar{\M}_{a}^{(2)}V_\phi \M_b V_{\phi^\dagger} k_\mu\bar{\M}_{a}^{(1)\mu}\big]
=
\text{tr}\hsm\big[\M_b V_{\phi^\dagger} k_\mu\bar{\M}_{a}^{\mu}V_\phi\big]
+ (\text{terms}\hs\propto\epsilon)\hs, ~~~
\end{equation}
where the block $\bar{\M}_{a}^{\mu}\!=\!\bar{\M}_{a}^{(1)\mu}\bar{\M}_{a}^{(2)}$ 
is now separated into two parts located at the beginning and end of the trace, 
and the extra terms proportional to $\epsilon\hs$ arise from the non-cyclic trace of NDR.\ 
Since the proof of the Ward-Takahashi identity \eqref{eq:WT_QED} involves propagators  
in both $\bar{\M}_{a}^{(1)\mu}$ and $\bar{\M}_{a}^{(2)}\hsm$, one needs to move $\bar{\M}_{a}^{(2)}$ 
to the end of the trace to complete the block $\bar{\M}_{a}^{\mu}\hs$, which leads to 
extra terms proportional to $\epsilon$ that then combine with the $1/\epsilon$ divergence 
of the hard part to produce a finite contribution, violating the identity \eqref{eq:WT_momentum_space}.\ 
Hence, only the Yukawa vertices can be chosen as the reading point.\
The above argument for a Yukawa vertex as reading point can be readily 
generalized to an arbitrary number of Yukawa vertices at the one-loop level.

\vspace*{0.8mm}

It was also suggested in the literature\,\cite{Kreimer:1993bh} 
not to choose the gauge vertices as reading
points in order to maintain recursive renormalizability of the full result of the diagrams
as well as its gauge invariance.\ In the above analysis, we support this rule for a
very different reason, namely, the correspondence of the hard part
in the method of region to a one-loop gauge-invariant EFT operator requires
choosing the Yukawa vertex as the reading point. 

\vspace*{0.8mm}

This choice of reading point is also convenient when performing the
matching procedure by using the covariant derivative expansion (CDE)\cite{Henning:2014wua}.%
\footnote{%
Universal one-loop effective actions induced by heavy fermion loops 
were studied in the literature using the covariant derivative expansion up to Dim-6 operators\cite{Kramer:2019fwz,Ellis:2020ivx,Cohen:2020fcu}.\,
}\ 
In this method,
the one-loop effective nTGC operators are obtained by evaluating the
functional trace:
\begin{equation}
-\frac{\text{i}}{2}\text{STr}\!\(\frac{1}{K_{i}}U_{H^{\dagger}}^{ij}\frac{1}{K_{j}}U_{H}^{ji}+\cdots\)
\!,
\end{equation}
where, following the notation of~\cite{Cohen:2020fcu},
we split the block-diagonal interaction matrix $U$ into
$U_{H}^{}$ and $U_{H^{\dagger}}^{}$ corresponding to the type of Yukawa interaction, and
$K^{-1}$ is the propagator matrix.\  Following the previous argument
for explicit loop calculation, we have moved $U_{H}$ to the end of the trace, 
since it is the reading point.\footnote{The application of NDR and the subtlety of the choice of 
reading point when using the CDE were discussed by \cite{Fuentes-Martin:2022vvu} 
in the context of evanescent operators.}\ 
The $\gamma_{5}$ matrices are then moved to the right end of the
trace by commutation relations, followed by the replacement \eqref{eq:NDR_G5_replacement}
according to the NDR manipulation.\  
Using the public code {\tt STrEAM}~\cite{Cohen:2020qvb}, we have checked
that, after eliminating redundant operators, the CDE gives the same results
for the one-loop effective operator matching to the hard part.\ To obtain
the full vertex for the ``heavy-light'' case, we need further to 
compute the soft part by evaluating the loop contribution of the tree-level effective
operators using the same Higgs vertices as reading points and adding it
to the contribution of the one-loop effective operators.\  In this way,
irrelevant anomalies that usually appear in the EFT loop calculations
would never appear in any intermediate step of computing the 
full nTGC vertex.

\section{\large\hspace*{-2.5mm}Results for Induced nTGCs}
\label{sec:results}
\label{sec:5}

In this section, we present the results of loop calculations derived 
using the method of Section\,\ref{sec:method}.\  These results are
then combined with various nTGC vertices and matched to the one-loop effective
nTGC operators as given in Eqs.\eqref{eq:operators1}-\eqref{eq:operators2}.\ 
In our convention, the Higgs expectation value is given by  
$\langle\phi^{0}\rangle\!=\hsm v/\hsm\sqrt{2\,}$, where $\phi^{0}$ denotes
the neutral component of the SM Higgs doublet field $H\hs$.\

\subsection{\hspace*{-3mm}Heavy Fermion Loop Contributions to nTGCs}
\label{subsec:heavy_loop_result}
\label{sec:5.1}

We start with the simpler case, in which the nTGC vertices are
generated by one-loop contributions of the heavy fermions, including an SU(2) doublet
${\cal N}$ and a fermionic singlet ${\cal E}$ with hypercharges $Y_{{\cal N}}^{}$
and $Y_{{\cal E}}^{}\!=\!Y_{{\cal N}} -1/2$ that play the role of the fields
$N$ and $E$ in Eq.\eqref{eq:general_model}, respectively. 
We assume that these heavy fermions have nearly degenerate masses
$M_{{\cal N}}\approx M_{{\cal E}}\approx M$, 
so there is only one heavy mass scale for EFT matching.\  
The relevant Lagrangian terms take the following form: 
\begin{align}
{\cal L} & \,\supset\,\bar{{\cal N}}(\text{i}\slashed{D}\!-\!M_{{\cal N}}^{}){\cal N}
\!+\hsm \bar{{\cal E}}(\text{i}\slashed{D}\!-\!M_{{\cal E}}^{}){\cal E} 
\!+\!\bar{{\cal N}}H(c_{V}^{}\!+\!c_{A}^{}\gamma_{5}^{}){\cal E}\!+\hsm\text{h.c.} 
\label{eq:heavy_model}
\end{align}
In the cases of $(Y_{{\cal N}}^{},\,Y_{{\cal E}}^{})=(-1/2,-1)$,
$\left(1/2,0\right)$, and $\left(3/2,1\right)$, at least one of
the heavy fermions can mix with SM leptons through Yukawa
couplings to the Higgs doublet.\ In this subsection, we set these
heavy-light mixing couplings be negligibly small as compared to the
couplings between the heavy particles, and leave their treatment to
the next subsection.\ The absence of the heavy-light couplings can be ensured
by imposing a $Z_{2}^{}$ symmetry.\ 
The result for vertices and Wilson coefficients in this and next subsections 
are additive when a model generates both ``all heavy'' and ``heavy-light''
loop diagrams. 

\vspace*{0.8mm}

For the four types of basic one-loop diagrams of triple neutral gauge bosons 
in Fig.\,\ref{fig:basic_diagrams}, 
we compute the off-shell expressions from Eq.\eqref{eq:base_loop}, 
with the substitutions $N\!\to\!{\cal N}$
and $E\!\to\!{\cal E}$.\  Thus, we derive the following:
\begin{subequations}
\begin{align}
\Gamma_{1} & =\frac{\text{i}\hs c_{VA}^{}}{~240\pi^{2}M^{4}~}
\left[(4q^{2}\!+\!3p^{2}\!+\!4p\hsm\cdot\hsm q)q_{\sigma}^{}\epsilon^{\mu\nu\rho\sigma}
\!+\! (q\!\leftrightarrow\! p,\hs \nu\!\leftrightarrow\rho)\right]\!,
\\[0.6mm]
\Gamma_{2} & =\frac{\text{i}\hs c_{VA}^{}}{~240\pi^{2}M^{4}~}
\!\left[2(k^{\rho}\!-\!k^{\mu}\!+\!q^{\mu})k_{\alpha}^{}q_{\beta}^{}
\epsilon^{\nu\rho\alpha\beta}\!+\!(3k^{2}\hsm\!+\!q^{2}\hsm\!+\!4k\hsm\cdot\hsm q)
k_{\sigma}\epsilon^{\mu\nu\rho\sigma}
\!\!+\! (q\!\leftrightarrow\! k,\hs \nu\!\leftrightarrow\!\mu) \right]\! ,
\\[0.6mm]
\Gamma_{3} & =\Gamma_{1}\Big|_{c_{V}^{}\rightarrow c_{V}^{*},c_{A}\rightarrow-c_{A}^{*}},
\hspace*{4mm}
\Gamma_{4} =\Gamma_{2}\Big|_{c_{V}^{}\rightarrow c_{V}^{*},c_{A}\rightarrow-c_{A}^{*}},
\end{align}
\end{subequations}
with the coupling coefficient  
$c_{VA}^{2}\!=\! c_{V}^{}c_{A}^{*}+c_{A}^{}c_{V}^{*}$.\ 
Including the charges and gauge couplings corresponding to the different sets of the external gauge
bosons in Fig.\,\ref{fig:2} and using the Schouten
identity \eqref{eq:schouten_pp}, we can match these results
directly to the sets of operators in Eqs.\eqref{eq:operators1}-\eqref{eq:operators2}, 
since the loop integrals have no soft parts.\ 
The effective Lagrangian takes the following form:  
\begin{equation}
{\cal L}~\supset\,\sum_{I}\!c_{I}^{}\mathcal{O}_{I}^{}+\text{h.c.} \,,
\end{equation}
where the label $I$ runs over the labels of the nTGC operators $\mathcal{O}_{I}^{}$.\ 
We compute the one-loop Wlison coefficients $c_{I}^{}$ as follows:
\\[-8mm]
\begin{subequations}
\begin{align}
c_{\tilde{W}W}^{}  &= -\frac{g^{2}c_{VA}^{2}}{~240\pi^{2}M^{4}~} \,, 
\\
c_{\tilde{W}W}^{\prime} &= \frac{g^{2}c_{VA}^{2}}{~160\pi^{2}M^{4}~} \,, 
\\
c_{\tilde{B}B}^{} &= 
-\frac{~g'^{2}(1\!-\!5Y_{\cal N}^{}\!+\!10Y_{\cal N}^{2})c_{VA}^{2}~}
{960\pi^{2}M^{4}} \,, 
\\
c_{\tilde{B}B}^{\prime}  &= 
\frac{~g'^{2}(3\!-\!20Y_{\cal N}^{}\!+\!40Y_{\cal N}^{2})c_{VA}^{2}~}
{1920\pi^{2}M^{4}} \, , 
\end{align}
\label{all-heavy1}
\end{subequations}
 and 
\\[-11.5mm]
\begin{subequations}
\begin{align}
c_{\tilde{B}W}^{} &=  
-\frac{gg'c_{VA}^{2}}{~1920\pi^{2}M^{4}~} \,, 
\\
c_{\tilde{B}W}^{\prime} &= 
-\frac{~gg'(1\!-\!5Y_{\cal N}^{})c_{VA}^{2}~}{240\pi^{2}M^{4}} \,, 
\\
c_{\tilde{W}B}^{} &= 
\frac{~gg'(3\!-\!20Y_{\cal N}^{})c_{VA}^{2}~}{1920\pi^{2}M^{4}} \, , 
\end{align}
\label{all-heavy2}
\end{subequations}
\hspace*{-3mm}
where the coupling coefficients
$c_{V\!A}^{2}\!=\! c_{V}^{}c_{A}^{*}\!+\!c_{A}^{}c_{V}^{*}\,$.\ 
Using Eq.\eqref{eq:on-shell_coefficients}, we translate these results 
into the following on-shell coefficients:
\begin{subequations}
\begin{align}
\hspace*{-3mm}
c_{\gamma^{*}ZZ}^{} & =
\frac{m_{Z}^{5}\hs c_{V\!A}}{~192\pi^{2}v M^{4}~}
\sin(2\theta_{W}^{})(2Y_{\cal N}^{}\!-\!1)
\big[(2Y_{\cal N}^{}\!-\!1)\cos(2\theta_{W}^{})\!-\!2Y_{\cal N}^{}\big],
\\
\hspace*{-3mm}
c_{Z^{*}ZZ}^{} & =
\frac{m_{Z}^{5}\hs c_{VA}}{~1920\pi^{2}v M^{4}~}
\big[5(2Y_{\cal N}^{}\!-\!1)^{2}\cos(4\theta_{W}^{})\!-\!40(2Y_{\cal N}^{}\!-\!1)
\cos(2\theta_{W}^{})\!+\!60Y_{\cal N}^{2}\!-\!20Y_{\cal N}^{}\!+\!7\big],
\\
\hspace*{-3mm}
c_{\gamma^{*}\gamma Z}^{} & =
\frac{m_{Z}^{5}c_{VA}}{~192\pi^{2}v M^{4}~}
\sin^{2}(2\theta_{W}^{})(2Y_{\cal N}^{}\!-\!1)^{2},
\\
\hspace*{-3mm}
c_{Z^{*}\gamma Z}^{} & =
\frac{m_{Z}^{5}c_{VA}}{~192\pi^{2}v M^{4}~}
\sin(2\theta_{W}^{})(2Y_{\cal N}^{}\!-\!1)
\big[(2Y_{\cal N}\!-\!1)\cos(2\theta_{W}^{})\!-\hsm 2Y_{\cal N}^{}\big] ,
\end{align}
\end{subequations}
where the nTGC coupling coefficients $(c_{V^*\gamma Z}^{},\hs c_{V^*ZZ}^{})$
are connected to the conventional notations\,\cite{Degrande:2013kka}\cite{Ellis:2022zdw}\cite{Ellis:2023ucy} 
via $(c_{V^*\gamma Z}^{},\hs c_{V^*ZZ}^{})\!=\!(e h_3^V,\hs e f_5^V)\hs$,
as shown in Eq.\eqref{eq:CVVZ-h3f5}.

In passing, for an estimate, 
we consider the future $e^+e^-$ colliders CEPC (250$\hs$GeV) and CLIC (3$\hs$TeV)
and a future $pp$ collider (100$\hs$TeV), with an integrated luminosity 
$(20,\hs 5,\hs 30)\,$ab$^{-1}$ respectively.\ 
According to the collider analyses\,\cite{Ellis:2023ucy}-\cite{Liu:2024tcz}, 
they can probe the form factors $(h_3^Z,\,h_3^\gamma)$ down to 
$h_{3}^{Z}\!<\!(1.4\!\times\!10^{-4}\!,\, 6.2\!\times\!10^{-5}\!,\, 3.0\!\times\!10^{-7})$,
and $h_{3}^{\gamma}\!<\!(4.9\!\times\!10^{-4}\!,\, 1.0\!\times\!10^{-4}\!,\, 3.5\!\times\!10^{-7})$, 
respectively, and we take just one
nTGC contribution at a time.\ 
For $Y_{\cal N}^{}\!=\!-\fr{1}{2}\hs$, these bounds correspond
to $M/|c_{VA}|^{1/4}\!\!<\!\!(80,\,240,\,368)\hs$GeV with the $h_{3}^{Z}$ constraints alone,
and become $M/|c_{VA}|^{1/4}\!<\!(150,\,480,\,770)\hs$GeV with the $h_{3}^{\gamma}$ constraints alone.\
These sensitivities are quite weak because such fermionic UV contributions are suppressed
by both the heavy mass factor $\propto M^{-4}$ and the one-loop factor.\  
A more careful phenomenological analysis is needed to extract
the actual sensitivity, including contributions of the interference between the
$Z^{*}$-exchange and $\gamma^{*}$-exchange channes.\
This will improve the sensitivity reaches on $M/|c_{VA}|^{1/4}$.\ 
These analyses are useful for the phenomenology of strongly-coupled UV models of new physics.\
In particular, discovery at the LHC or a future collider of an nTGC coupling in the absence of 
a new particle would be an indicator of a strongly-interacting sector beyond the SM.

\subsection{\hspace*{-3mm}nTGCs from Fermion Loops with Heavy-Light Mixing}
\label{subsec:Heavy_light_result}
\label{sec:5.2}

In this subsection, we extend our analysis to include one-loop contributions 
where the heavy and light fermions mix through a Yukawa-type coupling to the SM Higgs doublet.\ 
We begin by presenting general off-shell expressions for the one-loop diagrams of
triple neutral gauge bosons in Fig.\,\ref{fig:1}, setting 
$M_{N}^{} \!=\! M$, $M_{E}^{}\!=\!0$ and $V_{H}^{\pm}\!=\!(1\pm\gamma_{5}^{})/2\,$
in the propagators of Eq.\eqref{eq:base_loop}, 
identifying the fields $N$ and $E$ with heavy and light
fields respectively:\footnote{%
The results differ only by a minus sign for the opposite assignment of 
$V_{H}^{\pm}\!\to\!(1\mp\gamma_{5})/2\,$.}  
\begin{subequations}
\begin{align}
\hspace*{-3mm}
\Gamma_{1}^{h} & =
\frac{\text{i}}{\,12\pi^{2}M^{4}\,}
\bigg[\!\!\(\!-\Delta\!-\hsm\frac{11}{6}\hsm\)\hsm\!(p\hsm\cdot\hsm q)q_{\sigma}\epsilon^{\mu\nu\rho\sigma}
\!+\!\(\hsm\frac{1}{2}\Delta\!+\hsm\frac{5}{6}\)\!\!q^{2}p_{\sigma}\epsilon^{\mu\nu\rho\sigma}
\!+\!\hsm\(\!\Delta\!+\hsm\frac{25}{12}\)\hsm\!q^{\mu}p_{\alpha}q_{\beta}\epsilon^{\nu\rho\alpha\beta}
\nonumber \\
\hspace*{-3mm}
&\hspace*{10mm}
+\frac{1}{12}q^{\nu}p_{\alpha}q_{\beta}\epsilon^{\mu\rho\alpha\beta}
\!-\hsm\frac{1}{2}q^{\rho}p_{\alpha}q_{\beta}\hs\epsilon^{\mu\nu\alpha\beta}
\!\! +\hsm (q\!\leftrightarrow\! p,\hs \nu\!\leftrightarrow\!\rho)\hsm \bigg]\hsm,
\\
\hspace*{-3mm}
\Gamma_{1}^{} & =
\frac{\text{i}}{\,12\pi^{2}M^{4}\,}\bigg[\hsm\frac{1}{4}(p\hsm\cdot\hsm q)q_{\sigma}
\epsilon^{\mu\nu\rho\sigma}
\!+\!\hsm\(\frac{1}{2}\log\hsm\frac{M^{2}}{\hs -k^2\hs}
\!-\!\frac{5}{12}\hsm\)\hsm\!q^{2}p_{\sigma}^{}\epsilon^{\mu\nu\rho\sigma}
\nonumber \\
\hspace*{-3mm}
&\hspace*{10mm}
+\!\(\frac{1}{2}\!-\!\log\hsm\frac{M^{2}}{\hs -k^{2}\hs}\!\)\!q^{\nu}p_{\alpha}^{}q_{\beta}
\epsilon^{\mu\rho\alpha\beta}
\!+\!\(\!-\frac{11}{12}\!+\hsm\log\hsm\frac{M^{2}}{\hs -k^2\hs}\)\!
q^{\rho}p_{\alpha}q_{\beta}\hs\epsilon^{\mu\nu\alpha\beta}\! +\hsm 
(q\!\leftrightarrow\! p,\hs \nu\!\leftrightarrow\!\rho)\hsm \bigg]\hsm,
\\
\hspace*{-3mm}
\Gamma_{3}^{h} & =
\frac{\text{i}}{\,12\pi^{2}M^{4}\,}\bigg[\hsm\frac{19}{\hs 12\hs}(p\hsm\cdot\hsm q)
q_{\sigma}\epsilon^{\mu\nu\rho\sigma}
\!+\!\(\!\Delta\!-\hsm\frac{1}{3}\)\hsm\!q^{2}p_{\sigma}\epsilon^{\mu\nu\rho\sigma}
\!-\!\frac{7}{\hs 12\hs}q^{\mu}p_{\alpha}q_{\beta}\hs\epsilon^{\nu\rho\alpha\beta}\nonumber \\
\hspace*{-3mm} 
&\hspace*{10mm}
+\(\hsm\frac{11}{12}\!-\!\Delta\!\)\hsm\!q^{\nu}p_{\alpha}q_{\beta}\hs\epsilon^{\mu\rho\alpha\beta}
\!-\!\frac{9}{\hs 4\hs}q^{\rho}p_{\alpha}q_{\beta}\hs\epsilon^{\mu\nu\alpha\beta}\! +\hsm
(q\!\leftrightarrow\! p,\hs \nu\!\leftrightarrow\!\rho)\hsm \bigg]\hsm,
\\
%
\hspace*{-3mm}
\Gamma_{3}^{} & =
\frac{\text{i}}{\,12\pi^{2}M^{4}\,}
\bigg[\hsm\frac{19}{\hs 12\hs}(p\hsm\cdot\hsm q)q_{\sigma}\hs\epsilon^{\mu\nu\rho\sigma}
\!-\!\(\!2\!+\!\log\hsm\frac{M^{2}}{\hs -q^{2}\hs}\)\!q^{2}p_{\sigma}^{}\hs\epsilon^{\mu\nu\rho\sigma}
\!-\!\frac{7}{\hs 12\hs}\hs q^{\mu}p_{\alpha}^{}q_{\beta}^{}\hs\epsilon^{\nu\rho\alpha\beta}
\nonumber 
\\
\hspace*{-3mm}
&\hspace*{10mm} +\!\(\!\frac{31}{\hs 12\hs}\!+\!\log\hsm\frac{M^{2}}{\hs -q^{2}\hs}\)\!
q^{\nu}p_{\alpha}^{}q_{\beta}^{}\hs\epsilon^{\mu\rho\alpha\beta}
\!-\!\frac{9}{\hs 4\hs}q^{\rho}p_{\alpha}^{}q_{\beta}^{}\hs\epsilon^{\mu\nu\alpha\beta}\! 
+\hsm (q\!\leftrightarrow\! p,\hs \nu\!\leftrightarrow\!\rho)\hsm \bigg]\hsm,
\end{align}
\label{eq:general_heavy_light_loop1}
\end{subequations}
and
\begin{subequations}
\begin{align}
\hspace*{-3mm}
\Gamma_{2}^{h} & =
\frac{\text{i}}{\,96\pi^{2}M^{4}\,}
\bigg[\hsm -\frac{5}{\hs 3\hs} (k\hsm\cdot\hsm q)
q_{\sigma}^{}\hs\epsilon^{\mu\nu\rho\sigma}\!+\hsm\frac{1}{3}q^{2}k_{\sigma}\hs\epsilon^{\mu\nu\rho\sigma}
\!+\!q^{\mu}k_{\alpha}q_{\beta}^{}\hs\epsilon^{\nu\rho\alpha\beta}
\nonumber \\
&
\hspace*{-3mm}  
\hspace*{10mm}
+q^{\nu}k_{\alpha}^{}q_{\beta}^{}\hs\epsilon^{\mu\rho\alpha\beta}
\!-\!\frac{5}{\hs 3\hs}q^{\rho}k_{\alpha}^{}q_{\beta}^{}\hs\epsilon^{\mu\nu\alpha\beta}\! 
+\hsm (q\!\leftrightarrow\! k,\hs \nu\!\leftrightarrow\!\mu)\hsm \bigg]\hsm,
\\
\hspace*{-3mm}
\Gamma_{2} & =\Gamma_{2}^{h}\,,
\\
\hspace*{-3mm}
\Gamma_{4}^{h} & 
=\frac{\text{i}}{\,24\pi^{2}M^{4}\,}\hsm
\bigg[\!\!\(\!2\Delta\!+\!\frac{8}{\hs 3\hs}\hsm\)\!
(k\hsm\cdot\hsm q)q_{\sigma}^{}\hs\epsilon^{\mu\nu\rho\sigma}
\!-\!\(\!\Delta\!+\!\frac{\hs 11\hs}{6}\)\hsm\!
q^{2}k_{\sigma}\hs\epsilon^{\mu\nu\rho\sigma}
\!+\hsm q^{\mu}k_{\alpha}^{}q_{\beta}^{}\hs\epsilon^{\nu\rho\alpha\beta}
\nonumber 
\\
\hspace*{-3mm} 
& \hspace*{10mm} 
+q^{\nu}k_{\alpha}^{}q_{\beta}^{}\hs\epsilon^{\mu\rho\alpha\beta}
\!-\!\(\!2\Delta\!+\!\frac{8}{\hs 3\hs}\)\hsm\!q^{\rho}k_{\alpha}^{}q_{\beta}^{}\hs\epsilon^{\mu\nu\alpha\beta}
\!+\hsm (q\!\leftrightarrow\! k,\hs \nu\!\leftrightarrow\!\mu)\hsm \bigg]\hsm,
\\
\hspace*{-3mm}
\Gamma_{4}^{} & =\frac{\text{i}}{\,12\pi^{2}M^{4}\,}
\bigg[\!\!\(\!\frac{1}{\hs 6\hs}\!-\hsm\log\hsm\frac{M^{2}}{\hs -p^{2}\hs}\!\)\hsm\!
(k\hsm\cdot\hsm q) q_{\sigma}^{}\hs\epsilon^{\mu\nu\rho\sigma}
\!-\! \(\!\frac{7}{\hs 12\hs}\!-\!\frac{1}{\hs 2\hs}\log\hsm\frac{M^{2}}{\hs -p^{2}\hs}\)\!q^{2}k_{\sigma}^{}\hs\epsilon^{\mu\nu\rho\sigma}
\nonumber 
\\
\hspace*{-3mm}
&
\hspace*{10mm} +q^{\nu}k_{\alpha}^{}q_{\beta}^{}\hs\epsilon^{\mu\rho\alpha\beta}
-\!\(\!\frac{1}{\hs 6\hs}
\!+\hsm\log\hsm\frac{M^{2}}{\hs -p^{2}\hs}\!\)\!
q^{\rho}k_{\alpha}^{}q_{\beta}^{}\hs\epsilon^{\mu\nu\alpha\beta}\!+\hsm 
(q\!\leftrightarrow\! k,\hs \nu\!\leftrightarrow\!\mu)\hsm \bigg]\hsm,
\end{align}
\label{eq:general_heavy_light_loop2}
\end{subequations}
\hspace*{-3mm}
where $\Gamma_{i}^{h}$ ($\Gamma_{i}$) is the hard part and 
the full result (soft$+$hard) for each type of diagram in Fig.\,\ref{fig:basic_diagrams}, and
$\,\Delta\!=\! 1/\epsilon\!-\!\gamma_{E}^{}\!+\!\log(4\pi)\!+\!\log(\mu^{2}\hsm/M^{2})\hs$.\ 
We see that the divergence and the logarithmic renormalization scale dependence 
cancels correctly between the soft and hard parts.\ 
The remaining logarithmic factors take the form of 
$\log\hsm\frac{M^{2}}{\hs -Q^{2}\hs}\hs$, 
where $Q\,(=\!k, p, q)$ is one of the external momenta, 
and describes the IR divergence of the loop diagram as $\hs Q\!\to\!0\hs$.\

\vspace*{0.8mm}

The results for specific models with heavy-light mixing loops can be
obtained by inserting the corresponding gauge couplings into Eqs.\eqref{eq:general_heavy_light_loop1}
and \eqref{eq:general_heavy_light_loop2}.\ 
As a concrete example, we consider an extension
of the SM with a weak SU(2) fermion doublet 
$F\!=\!(f^{0},f^{-})^{T}$ with hypercharge 
$Y_{F}\!=\!-\fr{1}{2}\hs$.\ Thus, the relevant new physics Lagrangian reads,
\begin{align}
{\cal L} \,\supset\, \bar{F}(\text{i}\slashed{D}\!-\!M)F\!+\hsm 
(y\bar{F}H e_{R}^{}\hsm +\hsm\text{h.c.}) \hs.
\label{eq:heavy_light_Y1/2}
\end{align}
The mixing mass term $\mu_{FL}^{}\bar{L}F+\text{h.c.}$ between the heavy fermion
and the SM left-handed lepton doublet $L$ can be eliminated by a field
redefinition, so Eq.\eqref{eq:heavy_light_Y1/2} presents the Lagrangian terms 
after this redefinition.\ 
With these, we derive the one-loop effective coefficients of 
the nTGC operators \eqref{eq:operators1} and \eqref{eq:operators2} as follows:
\begin{subequations}
\begin{alignat}{3}
c_{\tilde{W}W}^{} &= -\frac{g^{2}y^{2}}{\,192\pi^{2}M^{4}\,}\,,
~~~~~~&
c_{\tilde{W}W}^{\prime} &= \frac{g^{2}y^{2}}{\,144\pi^{2}M^{4}\,}\,,
\\
c_{\tilde{B}B}^{} &= \frac{11g^{\prime\hs 2}y^{2}}{\,768\pi^{2}M^{4}\,}\,,
~~~~~~&
c_{\tilde{B}B}^{\prime} &= \frac{g^{\prime\hs 2}y^{2}}{\,576\pi^{2}M^{4}\,}\hsm\!
\(\hsm\!1\!+\hsm 6\log\hsm\frac{\mu^{2}}{M^{2}}\!\) \!,
\end{alignat}
\label{heavy-light1}
\end{subequations}
 and 
 \\[-10mm]
\begin{subequations}
\begin{align}
c_{\tilde{B}W}^{} &= 
\frac{g'gy^{2}}{\,1152\pi^{2}M^{4}\,}\!\(\!35\!+\!12\log\hsm\frac{\mu^{2}}{\hs M^{2}\hs}\!\)\!,
\\
c_{\tilde{B}W}^{\prime} &= 
\frac{g'gy^{2}}{\,144\pi^{2}M^{4}\,}\!\(\!4\!+\!3\log\hsm\frac{\mu^{2}}{M^{2}}\!\)
\!,
\\
c_{\tilde{W}B}^{} &= 
-\frac{g'gy^{2}}{\,1152\pi^{2}M^{4}\,}\!
\(\!17\!+\!12\log\hsm\frac{\mu^{2}}{\hs M^{2}\hs}\!\)
\!.
\end{align}
\label{heavy-light2}
\end{subequations}
\hspace*{-3mm}
These coefficients correspond to the hard parts of one-loop diagrams and
have no physical significance by themselves alone, since one needs to also include the
soft parts of the loops as obtained by tree-level matching.\ 
The full loop-contributions  
to the on-shell vertices can be summarized in the following form:
\begin{subequations}
\begin{align}
\Gamma_{V^{*}\gamma Z}^{\mu\nu\alpha}(q,p_{1}^{},p_{2}^{}) 
& = \frac{\,c^\prime_{V^{*}\gamma Z}\,}{m_Z^2} (q^{2}\!-m_{V}^{2})\hs 
p_{1\beta}^{}\hs\epsilon^{\mu\nu\alpha\beta} ,
\\[0.7mm]
\Gamma_{V^{*}ZZ}^{\mu\nu\alpha} (q,p_{1}^{},p_{2}^{}) 
& = \frac{1}{\,m_Z^2\,}\!
\left[c^\prime_{V^{*}ZZ}(q^{2})q^{2}\!-c_{V^{*}ZZ}^\prime(m_{V}^{2})\hs m_{V}^{2}\right]\!
\!\(p_{1}^{}\!-p_{2}^{}\)_{\beta}^{}
\epsilon^{\mu\nu\alpha\beta} ,
\end{align}
\label{eq:on-shell_vertices2} 
\end{subequations}
with the effective coupling coefficients given by 
\begin{subequations}
\label{eq:c-V*VZ-HL}
\begin{align}
c^\prime_{\gamma^{*}ZZ}(q_{\gamma^{*}}) & =
\frac{m_{Z}^{5}\hs y^{2}}{\,288\pi^{2}v M^{4}\,}\sin(2\theta_{W}^{})
\!\!\left[\hsm -3\cos(2\theta_{W}^{})\!+\!1\!+\!6\log\hsm\frac{M^{2}}{\hs -q_{\gamma^{*}}^{2}\hs}\right]\!,
\\
c^\prime_{Z^{*}ZZ}(q_{Z^{*}}) & =-\frac{m_{Z}^{5}\hs y^{2}}{\,576\pi^{2}v M^{4}\,}
\!\Bigg[\hsm 3\cos(4\theta_{W}^{})\!-\!20\cos(2\theta_{W}^{})
 \!+\!13\!+\!24\sin^{2}\!\theta_{W}\hsm\log\hsm\frac{M^{2}}{\hs -q_{Z^{*}\hs}^{2}\hs}\!\Bigg]\hsm,
\\
c^\prime_{\gamma^{*}\gamma Z} & =
-\frac{m_{Z}^{5}\hs y^{2}}{\,96\pi^{2}v M^{4}\,}\sin^{2}\hsm (2\theta_{W}^{})\hs,
\\
c^\prime_{Z^{*}\gamma Z} & 
=\frac{m_{Z}^{5}\hs y^{2}}{\,96\pi^{2}v M^{4}\,}
\sin(2\theta_{W}^{})\big[\!-\hsm\cos(2\theta_{W}^{})+3\hs\big]\hsm.
\end{align}
\end{subequations}
We note that Eq.\eqref{eq:on-shell_vertices2} is an extension of 
Eq.\eqref{eq:on-shell_vertices} to accommodate the logarithmic momentum dependence from the soft part.\ 
The nTGC coupling coefficients $(c'_{V^*\gamma Z},\hs c'_{V^*ZZ})$
are connected to the conventional notations via 
$(c'_{V^*\gamma Z},\hs c'_{V^*ZZ})\!=\!(e h_3^V,\hs e f_5^V)\hs$, as shown in Eq.\eqref{eq:CVVZ-h3f5}.

\vspace*{0.8mm}

For an estimate we consider the recent collider analyses\,\cite{Ellis:2023ucy}-\cite{Liu:2024tcz} 
on probing nTGCs at the future $e^+e^-$ colliders CEPC (250{\hs}GeV) and CLIC (3{\hs}TeV) 
and a future $pp$ collider (100{\hs}TeV), 
with integrated luminosityies $(20,\hs 5,\hs 30)\,$ab$^{-1}$ respectively.\ 
We find that the sensitivity reaches are
$M/|y|^{1/2}\!<\!(190,\,570,\,880)\hs$GeV for $Z^{*}$-exchange
and $M/|y|^{1/2}\!<\! (125,\,396,\,647)\hs$GeV
for $\gamma^{*}$-exchange.\
Since the fermionic UV contributions to nTGCs are suppressed
by both the heavy mass factor $\propto M^{-4}$ and the one-loop factor,   
the estimated collider bounds above and in Sec.\,\ref{sec:5.1} are quite weak.\ 
The bounds on the ``all-heavy'' and ``heavy-light'' cases are quite comparable to each other.\

During the finalization of this paper 
we compared our results with those of a recent paper\,\cite{Cepedello:2024ogz} that also
studied the derivation of nTGCs from certain fermionic UV models.\ 
This paper considered only 4 CP-even dimension-8 
operators in its Eqs.(2.4)-(2.7) that contribute to nTGC vertices 
with two on-shell gauge bosons, whereas our study considers a complete set
of 7 CP-conserving, Higgs-dependent dimension-8 operators 
\eqref{eq:operators1}-\eqref{eq:operators2} that generate nTGCs
and studies their matching to the one-loop contributions of UV models.\ 
These operators all contribute to the off-shell nTGC vertices and their consideration eliminates the possible ambiguity
that may arise from the choice of nTGC operator basis.\ 
Also, our method of matching differs from that of \cite{Cepedello:2024ogz}.\ 
Besides the coefficients of one-loop effective operators, we have provided a systematic treatment of 
the (off-shell) full nTGC vertices as obtained from the one-loop fermionic UV contributions, 
including both their hard parts and soft parts.\ The soft parts are induced by the heavy-light mixing case
and were not considered in \cite{Cepedello:2024ogz}.\
Our work provides an independent full treatment on the fermionic UV completion of the low energy nTGCs.\

\section{\large\hspace*{-2.5mm}Conclusions}
\label{sec:6}
\label{sec:conclusions}

Neutral triple gauge couplings (nTGCs) open up a unique window for probing the new physics beyond the Standard Model (SM), 
because they are absent both in the SM and in the SMEFT at the level of dimension-6 operators, 
and first appear in the SMEFT at the level of dimension-8 operators.\  
In recent years there has been increasing experimental and phenomenological interest  
in studying probes of neutral Triple Gauge Couplings (nTGCs) at present and future collider 
experiments\,\cite{ATLAS:2018nci}\cite{CMS:2016cbq}\cite{Ellis:2023ucy}-\cite{nTGC-other}.\ 
It is thus highly desirable to study how the underlying UV dynamics of new physics  
can naturally generate such nTGCs at low energy in the SMEFT formulation.\  

In this work, we have shown how nTGCs may be generated by loop diagrams involving vector-like heavy fermions, 
considering both loops of heavy fermions alone and also loops containing a mixture of the heavy fermions  
and the SM light fermions.\ 
We presented a complete set of 7 dimension-8 SMEFT operators \eqref{eq:operators1}-\eqref{eq:operators2} 
that generate CP-conserving off-shell nTGCs, 
where only 4 of them contribute to the nTGC form factors with two on-shell gauge bosons.\
Then, we demonstrated that at the one-loop order such a fermionic UV completion only induces the dimension-8 
nTGC operators containing two Higgs-doublet fields.

We have described the treatment of $\gamma_5$ in our fermionic one-loop analysis.\  
Then, we analyzed in detail the separation between the soft and hard parts of the one-loop integrals that appear 
in the heavy-light fermion mixing case and the associated Ward-Takahashi identity.\  
We further gave a prescription for the treatment of spinor traces that eliminates irrelevant anomalies 
in all the intermediate steps of matching.

We have evaluated the fermion loops with off-shell external gauge bosons and matched their
hard parts to the 7 dimension-8 CP-even nTGCs operators \eqref{eq:operators1}-\eqref{eq:operators2}.\ 
Then, we required two external gauge bosons of the nTGC vertices to be on-shell and derived the 4 form factors 
induced by the fermion loops.\ 
We have found that the contributions of the all-heavy and heavy-light fermion loops  
yield results of comparable magnitude, as can be seen by comparing Eqs.\eqref{all-heavy1}-\eqref{all-heavy2}) with Eqs.\eqref{heavy-light1}-\eqref{heavy-light2}.\ 
An essential difference is the appearance of logarithmic contributions 
in the heavy-light case that are absent in the all-heavy case.\ 
For the heavy-light case, we presented a generalized nTGC form factor formulation 
in Eq.\eqref{eq:on-shell_vertices2} and derived the corresponding form factor coefficients in Eq.\eqref{eq:c-V*VZ-HL},
which explicitly contain extra terms with logarithmic momentum dependence.\ 
This explains why the conventional nTGC form factor formulation \eqref{eq:on-shell_vertices}  
should be extended to the new Eq.\eqref{eq:on-shell_vertices2}
in the heavy-light case.

The perturbative one-loop fermionic UV contributions to the low-energy effective nTGC operators 
are suppressed by both the fourth power of the heavy fermion mass $M$ and a loop factor, making it
quite challenging to probe such perturbative scenarios of new physics via their UV contributions
to nTGCs at the LHC and future high energy colliders.\ On the other hand, the nTGCs may receive more sizeable 
contributions from certain strongly-interacting non-perturbative UV models.\
Thus, the possible collider discovery of a nTGC without an accompanying new particle
could provide evidence for a strongly-interacting sector beyond the SM.

\section*{\large Acknowledgments}
\vspace*{-2mm}
We thank Xiaochuan Lu, Ming-Lei Xiao and Jiang-Hao Yu for useful discussions.
The work of J.E. was supported in part by the United Kingdom STFC Grant ST/T000759/1.\
The work of HJH, RQX, SPZ and JZ was supported in part by the NSFC Grants 12175136 and 11835005.\
RQX has also been supported by an International Postdoctoral Exchange Fellowship.

\vspace*{10mm}

\appendix

\noindent
{\bf\Large Appendix:}

\section{\large\hspace*{-2.3mm}Off-Shell nTGC Vertices from Dimension-8 Operators}
\label{app:A}
\label{apd:off-shell_vertices}

In this Appendix, we show that among the nTGC operators
in Eqs.\eqref{eq:operators1} and \eqref{eq:operators2}, only four operators (combinations),
$\O_{\!\tilde{B}W}^{\prime}$, $\O_{\!\tilde{B}B}^{\prime}$,
$O_{\!\tilde{W}W}^{\prime}$, and $\O_{\!\tilde{B}W}^{}\!-\hsm\O_{\tilde{W}B}$, contribute
to the off-shell nTGC vertices that are phenomenologically relevant to fermion production processes at colliders.

For instance, we may consider the following production process:
\beq
\begin{array}{rl}
f\bar{f} \hspace*{-2.4mm} & \rightarrow V_{1}^{*}\rightarrow V_{2}^{*}V_{3}^{*},
\\[1mm]
V_{2}^{*} \hspace*{-2.4mm} & \rightarrow f_{2}\bar{f}_{2}\,, ~~~~~
V_{3}^{*} \rightarrow f_{3}\bar{f}_{3}\,,
\end{array}
\eeq
where the $V_{i}$ denotes neutral gauge bosons and the $f_{j}$ the SM fermions.\
For a nTGC vertex
$\Gamma_{V_{1}V_{2}V_{3}}^{\alpha_{1}\alpha_{2}\alpha_{3}}(p_{1},p_{2},p_{3})$
with momentum conservation $p_{1}^{}\!+p_{2}^{}\!+p_{3}^{}\!=\!0$,
we can ignore the terms that are proportional to $p_{i}^{\alpha_{i}}$,
because in the tree-level amplitude of the above production process,
each momentum $p_{i}^{\alpha_{i}}$ will be contracted with an external fermion current
and this contraction vanishes due to the equation of motion of the external on-shell fermions.

\vspace*{1mm}

Then, we derive the off-shell nTGC vertices from the operators \eqref{eq:operators1}-\eqref{eq:operators2}.\
For this, we will use the Schouten identity Eq.\eqref{eq:Schouten} to rearrange the expressions
such that only one momentum contracts with the antisymmetric tensor $\epsilon^{\mu\nu\al\beta}$.
Ignoring the terms that contain $p_{i}^{\alpha_{i}}$, the off-shell nTGC
vertices are evaluated as follows, where $\Gamma$ and $\Gamma'$ vertices are generated
by operators $\O$ and $\O'$ in Eqs.\eqref{eq:operators1} and \eqref{eq:operators2},
respectively.\
\\[2mm]
$\bullet$~nTGC Vertex $A^{\mu*}(p_{1})A^{\nu*}(p_{2})Z^{\rho*}(p_{3})$:
\begin{subequations}
	\begin{align}
		\Gamma_{\tilde{B}W}^{\prime\,\mu\nu\rho}(p_{1},p_{2},p_{3}) &
		=\frac{1}{\,4\,}ev^{2}\!
		\(p_{2}^{2}p_{1\sigma}^{}\epsilon^{\mu\nu\rho\sigma}-p_{1}^{2}p_{2\sigma}^{}\epsilon^{\mu\nu\rho\sigma}\)\!,
		\\
		\Gamma_{\tilde{B}B}^{\prime\,\mu\nu\rho}(p_{1},p_{2},p_{3}) &
		=\frac{1}{\,2\,}ev^{2}\cot\theta_{W}^{}
		\!\(p_{1}^{2}p_{2\sigma}^{}\epsilon^{\mu\nu\rho\sigma}-p_{2}^{2}p_{1\sigma}^{}\epsilon^{\mu\nu\rho\sigma}\)\!,
		\\
		\Gamma_{\tilde{W}W}^{\prime\,\mu\nu\rho}(p_{1},p_{2},p_{3}) &
		=\frac{1}{\,8\,}ev^{2}\tan\theta_{W}^{}\!
		\(p_{1}^{2}p_{2\sigma}^{}\epsilon^{\mu\nu\rho\sigma}-p_{2}^{2}p_{1\sigma}^{}\epsilon^{\mu\nu\rho\sigma}\)\!,
	\end{align}
\end{subequations}
which correspond to the contributions of $\O'_{\tilde{B}W}$, $\O'_{\tilde{B}B}$, and
$\O'_{\tilde{W}W}$ respectively, while the contributions by other operators vanish.\

\noindent
$\bullet$~nTGC Vertex $A^{\mu*}(p_{1})Z^{\nu*}(p_{2})Z^{\rho*}(p_{3})$:
\begin{subequations}
	\begin{align}
		\Gamma{}_{\tilde{B}W}^{\mu\nu\rho}(p_{1},p_{2},p_{3}) & =-\frac{1}{\,2\,}ev^{2}\csc2\theta_{W}\!\left[(p_{2}^{2}\!-\!p_{3}^{2})\hs p_{1\sigma}\epsilon^{\mu\nu\rho\sigma}\!+
		p_{1}^{2}(p_{2\sigma}^{}\!-\!p_{3\sigma}^{})\epsilon^{\mu\nu\rho\sigma}\right]\!,
		\\
		\Gamma{}_{\tilde{W}B}^{\mu\nu\rho}(p_{1},p_{2},p_{3}) &
		=\frac{1}{\,2\,}ev^{2}\csc2\theta_{W}\!
		\left[ (p_{2}^{2}-p_{3}^{2})\hs p_{1\sigma}^{}\epsilon^{\mu\nu\rho\sigma}
		\!+p_{1}^{2}(p_{2\sigma}-p_{3\sigma}^{})\epsilon^{\mu\nu\rho\sigma}\right]\!,
		\\
		\Gamma_{\tilde{B}W}^{\prime\,\mu\nu\rho}(p_{1},p_{2},p_{3}) & =\frac{1}{\,4\,}ev^{2}\tan\theta_{W}\!\left[\cot^{2}\theta_{W}(p_{2}^{2}\!-p_{3}^{2})\hs
		p_{1\sigma}^{}\epsilon^{\mu\nu\rho\sigma}
		\!+p_{1}^{2}(p_{2\sigma}^{}\!-p_{3\sigma}^{})\epsilon^{\mu\nu\rho\sigma}\right]\!,
		\\
		\Gamma_{\tilde{B}B}^{\prime\,\mu\nu\rho}(p_{1},p_{2},p_{3}) & =-\frac{1}{\,2\,}ev^{2}\!
		\left[(-p_{2}^{2}\!+\hsm p_{3}^{2})\hs p_{1\sigma}^{}\epsilon^{\mu\nu\rho\sigma}\!+
		p_{1}^{2}(p_{2\sigma}^{}\!-\hsm p_{3\sigma}^{})\epsilon^{\mu\nu\rho\sigma}\right]\!,
		\\
		\Gamma_{\tilde{W}W}^{\prime\,\mu\nu\rho}(p_{1},p_{2},p_{3}) & =\frac{1}{\,8\,}ev^{2}\!
		\left[(-p_{2}^{2}\!+\!p_{3}^{2})\hs p_{1\sigma}^{}\epsilon^{\mu\nu\rho\sigma}\!+
		p_{1}^{2}(p_{2\sigma}^{}\!-\hsm p_{3\sigma}^{})\epsilon^{\mu\nu\rho\sigma}\right]\!,
	\end{align}
\end{subequations}
which correspond to the contributions of $\O^{}_{\tilde{B}W}$, $\O^{}_{\tilde{W}B}$, $\O'_{\tilde{B}W}$,
$\O'_{\tilde{B}B}$, and $\O'_{\tilde{W}W}$ respectively.\
The contributions to $A^*Z^*Z^*$ vertex from other operators vanish.\
Since $\Gamma{}_{\tilde{B}W}^{\mu\nu\rho}+\Gamma{}_{\tilde{W}B}^{\mu\nu\rho}=0\hs$,
this means that the combination $\O_{\tilde{B}W}^{}+\O_{\tilde{W}B}^{}$ does not contribute
to the nTGC vertex $A^{*}Z^{*}Z^{*}$.

\vspace*{2.5mm}
\noindent
$\bullet$~nTGC Vertex $Z^{\mu*}(p_{1})Z^{\nu*}(p_{2})Z^{\rho*}(p_{3})$:
\begin{subequations}
	\begin{align}
		\Gamma_{\tilde{B}W}^{\prime\,\mu\nu\rho}(p_{1},p_{2},p_{3}) & =\frac{1}{\,4\,}ev^{2}\!
		\left[(p_{1}^{2}-2p_{2}^{2}\!+p_{3}^{2})\hs p_{1\sigma}^{}\epsilon^{\mu\nu\rho\sigma}\!
		+(2p_{1}^{2}\!-p_{2}^{2}-p_{3}^{2})\hs p_{2\sigma}\epsilon^{\mu\nu\rho\sigma}\right]\!,
		\\
		\Gamma_{\tilde{B}B}^{\prime\,\mu\nu\rho}(p_{1},p_{2},p_{3}) & =\frac{1}{\,2\,}ev^{2}\tan\theta_{W}\!
		\left[(p_{1}^{2}\!-2p_{2}^{2}+p_{3}^{2})\hs p_{1\sigma}\epsilon^{\mu\nu\rho\sigma}
		\!+(2p_{1}^{2}\!-p_{2}^{2}\!-p_{3}^{2})\hs p_{2\sigma}^{}\epsilon^{\mu\nu\rho\sigma}\right]\!,
		\\
		\Gamma_{\tilde{W}W}^{\prime\,\mu\nu\rho}(p_{1},p_{2},p_{3}) & =\frac{1}{\,8\,}ev^{2}\cot\theta_{W}\!
		\left[(p_{1}^{2}\!-2p_{2}^{2}\!+p_{3}^{2})\hs p_{1\sigma}^{}\epsilon^{\mu\nu\rho\sigma}\!+
		(2p_{1}^{2}\!-p_{2}^{2}\!-p_{3}^{2})\hs p_{2\sigma}\epsilon^{\mu\nu\rho\sigma}\right]\!,
	\end{align}
\end{subequations}
which correspond to the contributions of $\O'_{\tilde{B}W}$, $\O'_{\tilde{B}B}$, and
$\O'_{\tilde{W}W}$ respectively.\
The contributions to $Z^*Z^*Z^*$ vertex from other operators vanish.\
We find that all the nTGC operators do not contribute to the vertices $A^{*}A^{*}A^{*}$.\

\vspace*{1mm}

In summary, for the dimension-8 operators \eqref{eq:operators1}-\eqref{eq:operators2}
only four operators, $\O_{\!\tilde{B}W}^{\prime}$, $\O_{\!\tilde{B}B}^{\prime}$,
$\O_{\!\tilde{W}W}^{\prime}$ and $\O_{\!\tilde{B}W}^{}\!\!-\!\O_{\!\tilde{W}B}^{}$,
contribute to the off-shell nTGC vertices.

\end{document}